\documentclass[aps,preprint]{revtex4}

\usepackage{times}
\usepackage{graphicx}

 \def\dt{\Delta t} 
  
\def\Pab{-\sigma_{\alpha\beta}}

\begin{document}

\title{Steady-state hydrodynamic instabilities of active liquid crystals: 
Hybrid lattice Boltzmann simulations}


\author{D. Marenduzzo$^1$, E. Orlandini$^2$, M.E. Cates$^1$, 
J.M. Yeomans$^3$}

\affiliation{$^1$ SUPA, School of Physics, University of Edinburgh,
Mayfield Road, Edinburgh EH9 3JZ, Scotland \\
$^2$ Dipartimento di Fisica and Sezione INFN, Universita' di Padova, Via 
Marzolo 8, 35131 Padova, Italy \\
$^3$ The Rudolf Peierls Centre for Theoretical Physics, 1 Keble Road, Oxford 
OX1 3NP, England}

\begin{abstract}
We report hybrid lattice Boltzmann (HLB) simulations of the hydrodynamics of 
an active nematic liquid crystal sandwiched between confining walls 
with various anchoring conditions. We confirm the existence of a transition
between a passive phase and an active phase, in which there is spontaneous flow
in the steady state. This transition is attained for sufficiently ``extensile''
rods, in the case of flow-aligning liquid crystals, and for sufficiently
``contractile'' ones for flow-tumbling materials. In a quasi-1D geometry, deep in the active phase
of flow-aligning materials, our simulations give evidence of
hysteresis and history-dependent steady states, as well as of 
spontaneous banded flow. Flow-tumbling materials, in contrast, 
re-arrange themselves so that only the two boundary layers flow in steady
state.  Two-dimensional simulations, with periodic boundary conditions,
show additional instabilities, with the spontaneous flow appearing as
patterns made up of ``convection rolls''. These results demonstrate a remarkable richness (including dependence on anchoring conditions) in the steady-state phase behaviour of active materials, even in the absence of external forcing; they have no counterpart for passive nematics. Our HLB methodology, which combines 
lattice Boltzmann for momentum transport with a finite difference scheme for the order parameter dynamics, offers a robust and efficient method for probing the complex hydrodynamic behaviour of active nematics.
\end{abstract}

\maketitle

\section{Introduction}

{Active viscoelastic gels such as suspensions of active particles
and active liquid crystals are soft materials receiving increasing theoretical 
and experimental attention 
\cite{ramaswamy,ramaswamy2,kruse1,kruse2,joanny,EPL,liverpool,ramaswamy3,llopis,ignacio,bray,peter,beads,cytoskeleton,active_actin_myosin,surrey,nedelec,kierfeld,lubensky,activeLB}.
Such materials are called ``active'' \cite{nomenclature} because} they continuously burn energy, for example in the form of 
ATP, and this drives them out of thermodynamic equilibrium even when there is 
no external force.  
Activity imparts non-trivial 
physical properties. Perhaps the most striking is that 
spontaneous flow can exist in non-driven active materials 
\cite{ramaswamy,ramaswamy2,kruse1,kruse2,joanny,EPL,liverpool}, in sharp 
contrast to their passive liquid crystalline counterparts. Thus such 
materials, while always remaining active in a microscopic sense, can 
undergo a phase transition from a passive phase (where activity is 
macroscopically incoherent) to an active phase (exhibiting spontaneous flow).

Active materials are typically encountered in biological contexts
(although non-biological counterparts may also be realized, for instance with
vibrated granular rods \cite{ramaswamy3}). Examples
include suspensions of bacterial swimmers \cite{ramaswamy,llopis,ignacio}, 
cell extracts \cite{bray,peter}, self-propelled colloidal 
particles \cite{beads}, and cytoskeletal gels interacting 
with molecular motors, such as actomyosin solutions or microtubular 
networks in the presence of kinesin 
\cite{cytoskeleton,active_actin_myosin,surrey,nedelec,kierfeld}.
Activity leads to striking phenomena such as bacterial swarming, 
cytoplasmic streaming and elastotaxis \cite{ramaswamy}. Furthermore, 
many biological gels, such as actin and neurofilament networks, thicken when
sheared \cite{lubensky}. This is the opposite of the typical behaviour of 
viscous polymeric fluids such as molten plastics, which flow more easily as 
shear stress increases. Activity has been suggested to be
amongst the possible causes of this peculiar flow response 
\cite{ramaswamy,activeLB}.

In this paper we present a series of hybrid lattice Boltzmann simulations of 
the hydrodynamic equations of motion of an active nematic liquid
crystal. {Derivations of the continuum equations we use
are given in, e.g., Refs. \cite{ramaswamy,kruse1,liverpool}
and are not repeated here.  However we are aware of no numerical studies
of the equations (with the exception
of our previous work in \cite{activeLB}, which is a short report using
a different algorithm). These are the main focus of our work.}
Our model
considers a varying order parameter so that defects are automatically
incorporated, as is flow-induced or paranematic ordering. 
We show that, in the limit of a uniaxial active liquid crystal
with spatially uniform and temporally constant magnitude of order parameter
(we call this limiting case the ``Ericksen--Leslie'' model in analogy with the
terminology usually adopted for passive liquid crystals),
our model reduces to the equations considered
in Ref. \cite{EPL}. 
We then consider the specific case of a material that is 
sandwiched between two infinite parallel planes at which the director
field is anchored along a given direction. We first 
choose the anchoring to be 
along one of the directions in the plane (homogeneous anchoring),
and we then work out the case in which
there is different (conflicting) anchoring at the two boundary plates
(homogeneous at the top, and homeotropic, i.e. normal to the surface,
at the bottom).
When the anchoring is the same at both boundaries we find that there is
a phase transition \cite{note_noneq}
between a passive and an active phase when 
the ``activity'' $\zeta$, a parameter which measures the coupling between
pressure tensor and order parameter (see Section II for details), 
exceeds in absolute value a finite
threshold. For flow-aligning materials, the transition occurs for
sufficiently extensile rods; 
for tumbling materials it occurs for sufficiently contractile ones.
(Here ``extensile" means tending to propel fluid outwards along the long axis or molecular director ${\bf n}$, drawing it in radially on the midplane. while ``contractile" means the opposite \cite{ramaswamy}.)
Mixed boundary conditions, instead, 
lead to a zero activity threshold.

For homogeneous anchoring, we compare the numerical phase boundary to the one 
found in Ref. \cite{EPL} via a linear 
stability analysis, finding a good agreement. 
However, we show that 
the velocity profile found from the stability analysis is itself unstable
away from the phase boundary. 

 We also explore the nature of the 
solutions of the equations of motion (director and velocity field profiles) 
deep in the active phase, where we find that  
representative flow-tumbling and flow-aligning materials 
behave in a vastly different manner. 
The former can sustain a quasi-Poiseuille or banded 
flow, while spontaneous flow in the latter gets increasingly confined to a 
region close to the boundaries. 

Far from the phase boundary between the active and the passive phase there is
strong hysteresis, with multistable and history-dependent solutions.
These suggest that deep in the active phase the dynamics 
might be chaotic. 
It would be interesting to further explore the connections between the
active nematic hydrodynamics deep in the active phase and the rheochaotic
behaviour which selected passive liquid crystals display 
when they are subjected to an external forcing 
\cite{forest,ramaswamy_chaos,mike}. There may also be qualitative analogies
to the weakly turbulent viscoelastic flow discussed in \cite{alexander}.

Finally, we consider a quasi-2D case of a thin extensile flow-aligning
active liquid crystal film, wrapped on a cylindrical surface
(i.e. with periodic boundary conditions). Our simulations
shows that there are additional instabilities in this geometry.
Spontaneous flow this time appears as convection rolls, which,
deeper in the active phase, transiently increase in number 
and eventually split up leading to a highly
distorted flowing director field pattern.

{We close this introduction with some notes on nomenclature and wording. 
Firstly, an active gel is different from
a fluid which is driven out of equilibrium by an {\em external} shear or
heat flow, cases for which there is an important and vast literature
(see e.g. \cite{onuki1,onuki2}). In an active gel the driving is
{\em internal}, as, for instance, a bacterium uses up ATP to propel itself.

Secondly, at first glance our system shares some aspects with fluids 
which are driven out of equilibrium by a chemical reaction. There is a
significant literature on reaction-diffusion equations which lead 
to pattern formation
\cite{chem_noneq_book,chem_noneq_review,muthukumar}. 
Ultimately, our systems are chemically driven 
(e.g. via ATP hydrolysis), but
they differ from conventional reaction-diffusion systems in two ways. 
Firstly, the underlying fluid has liquid crystalline order
even in the passive state. Secondly, the activity
enters the equations of motion through a modification of the
stress tensor in the Navier-Stokes equations by a term which is non-potential
(i.e. it cannot be derived on the basis of any free energy). 
This makes the equations of active systems quite distinct from those addressed
by reaction-diffusion models.}

\section{Models and methods}

\subsection{Equations of motion}

We employ a Landau-de Gennes free energy ${\cal F}$,
whose density we name $f$, to describe the
equilibrium of the active liquid crystal (LC) in its passive phase
(i.e. when the activity parameters are switched off, see below).
This free energy density can be written as a sum of two terms.
The first is a bulk contribution,
\begin{eqnarray}
{f}_1=\frac{A_0}{2}(1 - \frac {\gamma} {3}) Q_{\alpha \beta}^2 -
          \frac {A_0 \gamma}{3} Q_{\alpha \beta}Q_{\beta
          \gamma}Q_{\gamma \alpha}
+ \frac {A_0 \gamma}{4} (Q_{\alpha \beta}^2)^2,
\label{eqBulkFree}
\end{eqnarray}
while the second is a distortion term, which we take in a (standard) one-constant approximation as \cite{degennes}
\begin{equation}\label{distorsion}
{f}_2=\frac{K}{2} \left(\partial_\gamma Q_{\alpha \beta}\right)^2.
\end{equation}
In the equations above , $A_0$ is a constant, $\gamma$ controls the
magnitude of order (it may be viewed as an effective
temperature or concentration for thermotropic and lyotropic
liquid crystals respectively), while
$K$ is an elastic constant. 
${f}=f_1+f_2$ is a standard free energy density to describe
passive nematic liquid crystals \cite{degennes}. 
Here and in what follows Greek indices
denote cartesian components and summation over repeated indices is implied.

The anchoring of the director field on the boundary surfaces
(Fig. 1) to a chosen director $\hat{n}^0$
is ensured by adding a surface term
\begin{eqnarray}
f_s & = & \frac{1}{2}W_0 (Q_{\alpha \beta}-Q_{\alpha \beta}^0)^2\\
Q_{\alpha \beta}^0 & = & S_0 (n_{\alpha}^0n_{\beta}^0-\delta_{\alpha\beta}/3)
\end{eqnarray}
The parameter $W_0$ controls the strength of the anchoring, while $S_0$
determines the degree of the surface order. If the surface order is equal to
the bulk order, $S_0$ should be taken equal to $q$, the order
parameter in the bulk ($3/2$ times the largest eigenvalue of the {\bf Q}
tensor). $W_0$ is large (strong anchoring) in what follows.

The equation of motion for {\bf Q} is taken to be \cite{beris,O92,O99}
\begin{equation}
(\partial_t+{\vec u}\cdot{\bf \nabla}){\bf Q}-{\bf S}({\bf W},{\bf
  Q})= \Gamma {\bf H}+\lambda {\bf Q}
\label{Qevolution}
\end{equation}
where $\Gamma$ is a collective rotational diffusion constant,
and $\lambda$ is an activity parameter of the liquid crystalline gel.
{The form of Eq. \ref{Qevolution} was suggested on the
basis of symmetry in Refs. \cite{ramaswamy,kruse1} and derived 
starting from an underlying microscopic model in Ref. \cite{liverpool}.}
The first term on the left-hand side of Eq. (\ref{Qevolution})
is the material derivative describing the usual time dependence of a
quantity advected by a fluid with velocity ${\vec u}$. This is
generalized for rod-like molecules by a second term
\begin{eqnarray}\label{S_definition}
{\bf S}({\bf W},{\bf Q})
& = &(\xi{\bf D}+{\bf \omega})({\bf Q}+{\bf I}/3)+ ({\bf Q}+
{\bf I}/3)(\xi{\bf D}-{\bf \omega}) \\ \nonumber
& - & 2\xi({\bf Q}+{\bf I}/3){\mbox{Tr}}({\bf Q}{\bf W})
\end{eqnarray}
where Tr denotes the tensorial trace, while
${\bf D}=({\bf W}+{\bf W}^T)/2$ and
${\bf \omega}=({\bf W}-{\bf W}^T)/2$
are the symmetric part and the anti-symmetric part respectively of the
velocity gradient tensor $W_{\alpha\beta}=\partial_\beta u_\alpha$.
The constant $\xi$ depends on the molecular
details of a given liquid crystal.
The first term on the right-hand side of Eq. (\ref{Qevolution})
describes the relaxation of the order parameter towards the minimum of
the free energy. The molecular field ${\bf H}$ which provides the 
force for this motion is given by
\begin{equation}
{\bf H}= -{\delta {\cal F} \over \delta {\bf Q}}+({\bf
    I}/3) Tr{\delta {\cal F} \over \delta {\bf Q}}.
\label{molecularfield}
\end{equation}

The fluid velocity, $\vec u$, obeys the continuity equation and
the Navier-Stokes equation,
\begin{equation}\label{navierstokes}
\rho(\partial_t+ u_\beta \partial_\beta)
u_\alpha = \partial_\beta (\Pi_{\alpha\beta})+
\eta \partial_\beta(\partial_\alpha
u_\beta + \partial_\beta u_\alpha)
\end{equation}
where $\rho$ is the fluid density, $\eta$ is an isotropic
viscosity, $\Pi_{\alpha\beta}=\Pi^{\rm passive}_{\alpha\beta}+
\Pi^{\rm active}_{\alpha\beta}$, and we have neglected an extra term
proportional to $\partial_{\alpha}u_{\alpha}$ which is zero in the
case we are interested in (incompressible fluids). The stress tensor
$\Pi^{\rm passive}_{\alpha\beta}$ necessary to describe ordinary LC
hydrodynamics is:
\begin{eqnarray}\label{BEstress}
\Pi^{\rm passive}_{\alpha\beta}= &-&P_0 \delta_{\alpha \beta} +2\xi
(Q_{\alpha\beta}+{1\over 3}\delta_{\alpha\beta})Q_{\gamma\epsilon}
H_{\gamma\epsilon}\\\nonumber
&-&\xi H_{\alpha\gamma}(Q_{\gamma\beta}+{1\over
  3}\delta_{\gamma\beta})-\xi (Q_{\alpha\gamma}+{1\over
  3}\delta_{\alpha\gamma})H_{\gamma\beta}\\ \nonumber
&-&\partial_\alpha Q_{\gamma\nu} {\delta
{\cal F}\over \delta\partial_\beta Q_{\gamma\nu}}
+Q_{\alpha \gamma} H_{\gamma \beta} -H_{\alpha
 \gamma}Q_{\gamma \beta} \\ \nonumber
& \equiv & \sigma_{\alpha\beta}+\tau_{\alpha\beta}
-\partial_\alpha Q_{\gamma\nu} {\delta
{\cal F}\over \delta\partial_\beta Q_{\gamma\nu}}.
\end{eqnarray}
In Eq. (\ref{BEstress}) we have defined the symmetric and 
anti-symmetric part of the passive stress tensor (not including the
double gradient term $\partial_\alpha Q_{\gamma\nu} {\delta
{\cal F}\over \delta\partial_\beta Q_{\gamma\nu}}$) as
$\sigma_{\alpha\beta}$ and $\tau_{\alpha\beta}$ respectively,
for later convenience.
$P_0$ is a constant in the simulations reported here.
The active term is given by
\begin{equation}
\Pi^{\rm active}_{\alpha\beta}=-\zeta  Q_{\alpha\beta}
\end{equation}
where $\zeta$ is a second activity constant \cite{ramaswamy,EPL}.
Note that with the sign convention chosen here $\zeta>0$  
corresponds to extensile rods and $\zeta<0$ to
contractile ones \cite{ramaswamy}.
{As for Eq. \ref{Qevolution}, the explicit form of the
active contribution to the stress tensor entering Eq. \ref{navierstokes}
was proposed on the basis of a symmetry analysis of a fluid
of contractile or extensile dipolar objects in \cite{ramaswamy}. It was
also derived by coarse graining a more microscopic model for a solution
of actin fibers and myosins in Ref. \cite{liverpool}.} 

A full understanding {of the physical origin (in both
bacterial suspensions and actomyosin gels)} of the
phenomenological couplings $\zeta$ and $\lambda$, as well
as of the range of values these may attain in physically relevant
situations, will require multi-scale modelling at different coarse 
graining levels, and more accurate quantitative experiments.
These are at the moment still lacking. However, we 
already know from experiments and from some
more microscopic approaches, that actomyosin gels are
contractile, so that in physiological conditions those materials
should be described by negative values of $\zeta$ \cite{thoumine}. The 
term proportional to $\lambda$ has been proposed in Ref. 
\cite{ramaswamy} as a symmetry allowed term which, for
dilute bacterial suspensions, should be negative and proportional to
the inverse of the time scale for relaxation of 
activity-induced ordering. In Ref. \cite{EPL} it was pointed out that, instead,
$\lambda>0$ when describing concentrated actomyosin gels and other
systems which display zipping or other
self-alignment effects (this is relevant for the cases considered in
\cite{frey}).

It is important to note that 
the model we have just written down
reduces for $\lambda=\zeta=0$ to the Beris-Edwards model for LC
hydrodynamics. For a sample
of uniaxial active LCs with a spatially uniform
degree of orientational order, the
director field (also called polarisation field in Refs. 
\cite{kruse1,kruse2,joanny})
 $\vec n$ is defined through
\begin{equation}\label{uniaxial}
Q_{\alpha\beta}=q\left(n_{\alpha}
n_{\beta}-\delta_{\alpha\beta}/3\right), 
\end{equation}
where $q$ is 
the degree of ordering in the system (assumed to be spatially uniform).
In this limit our model can be
shown to reduce to the vectorial model considered in
\cite{kruse1,joanny}, as will be shown explicitly in Section III.

\subsection{Hybrid lattice Boltzmann algorithm}

The differential equations (\ref{Qevolution}) and (\ref{navierstokes})
may both be solved by using a lattice Boltzmann (LB) algorithm \cite{succi}, 
based on
the 3-dimensional lattice Boltzmann algorithm for conventional liquid 
crystals \cite{lblc}, generalised to include the two extra active terms, 
as we discussed in Ref. \cite{activeLB}.

Here we use a different route, and solve Eq. (\ref{Qevolution})
via a finite difference predictor-corrector algorithm, while lattice Boltzmann
is used to solve the Navier-Stokes equation, (\ref{navierstokes}).
With respect to a full LB approach \cite{colin,lblc}, 
the primary advantage of this method is that it will allow simulations of 
larger systems as it involves
consistently smaller memory requirements. Indeed, while
in a full LB treatment one has to store 6 sets of 15 distribution functions at 
any lattice point
(if we choose the 3DQ15 velocity vector lattice \cite{succi}
as we do here), just 
one set of distribution functions plus the five independent components of
the ${\bf Q}$ tensor, is needed in this hybrid algorithm. 
Furthermore, we avoid in this
way the error term arising in the Chapman-Enskog expansion used to
connect the LB model to the order parameter evolution equation 
in the continuum limit \cite{colin}.

Lattice Boltzmann algorithms to solve the Navier-Stokes
equations of a simple fluid are defined in terms of a single set of partial
distribution functions, the scalars $f_i (\vec{x})$, that sum on each lattice
site $\vec{x}$ to give the density.
 Each $f_i$ is associated with a lattice
vector ${\vec e}_i$ \cite{lblc}.  We choose a 15-velocity model on the cubic
lattice with lattice vectors:
\begin{eqnarray}
\vec {e}_{i}^{(0)}&=& (0,0,0)\\
\vec {e}_{i}^{(1)}&=&(\pm 1,0,0),(0,\pm 1,0), (0,0,\pm 1)\\
\vec {e}_{i}^{(2)}&=&(\pm 1, \pm 1, \pm 1).
\label{latvects}
\end{eqnarray}
The indices, $i$, are ordered so that $i=0$ corresponds to
$\vec {e}_{i}^{(0)}$, $i=1,\cdots, 6$ correspond to the $\vec {e}_{i}^{(1)}$
set and $i=7,\cdots,14$ to the $\vec {e}_{i}^{(2)}$ set.
 For our hybrid code,
the input to the equilibrium distribution functions has to come from
the solution (via finite difference methods) of the coupled Eq.
(\ref{Qevolution}). This differs from the fully LB treatment of nematics; see
Refs. \cite{colin,lblc}.

Physical variables are defined as moments of the distribution functions:
\begin{equation}
\rho=\sum_i f_i, \qquad \rho u_\alpha = \sum_i f_i  e_{i\alpha}.
\label{eq1}
\end{equation}
The distribution functions evolve in a time step $\Delta t$ according
to
\begin{equation}
f_i({\vec x}+{\vec e}_i \Delta t,t+\Delta t)-f_i({\vec x},t)=
\frac{\Delta t}{2} \left[{\cal C}_{fi}({\vec x},t,\left\{f_i
\right\})+ {\cal C}_{fi}({\vec x}+{\vec e}_i \Delta
t,t+\Delta
t,\left\{f_i^*\right\})\right].
\label{eq2}
\end{equation}

This represents free streaming with velocity ${\vec e}_i$ followed by a
collision step which allows the distributions to relax towards
equilibrium.
The $f_i^*$'s are first order approximations to
$f_i({\vec {x}}+{\vec {e}}_i \dt,t+\dt)$, and 
they are obtained by using $\Delta t\, {\cal C}_{{f}i}({\vec
x},t,\left\{{f}_i \right\})$ on the right hand side of Eq. (\ref{eq2}).
Discretizing in this way, which is similar to a predictor-corrector
scheme, has the advantages that lattice viscosity terms are eliminated
to second order and that the stability of the scheme is improved \cite{colin}.

The collision operators are taken to have the form of a single
relaxation time Boltzmann equation, together with a forcing term
\begin{equation}
{\cal C}_{fi}({\vec {x}},t,\left\{f_i \right\})=
-\frac{1}{\tau_f}(f_i({\vec {x}},t)-f_i^{eq}({\vec {x}},t,\left\{f_i
\right\}))
+p_i({\vec {x}},t,\left\{f_i \right\}),
\label{eq4}
\end{equation}
The form of the equations of motion
follow from the choice of the moments of the equilibrium distributions
$f^{eq}_i$ and the driving terms $p_i$. Moreover, $f_i^{eq}$ is constrained by
\begin{equation}
\sum_i f_i^{eq} = \rho,\qquad \sum_i f_i^{eq} e_{i \alpha} = \rho
u_{\alpha}, \qquad
\sum_i f_i^{eq} e_{i\alpha}e_{i\beta} = \Pab+\rho
u_\alpha u_\beta
\label{eq6}
\end{equation}
where the zeroth and first moments are chosen to impose conservation of
mass and momentum. The second moment of $f^{eq}$ is determined
by $\sigma_{\alpha\beta}$, whereas the divergences of $\tau_{\alpha\beta}$
and of $\partial_\alpha Q_{\gamma\nu} {\delta
{\cal F}\over \delta\partial_\beta Q_{\gamma\nu}}$ enter effectively as 
a body force:
\begin{equation}
\sum_i p_i = 0, \quad \sum_i p_i e_{i\alpha} = \partial_\beta
\tau_{\alpha\beta}-\partial_\beta \left(\partial_\alpha Q_{\gamma\nu} {\delta
{\cal F}\over \delta\partial_\beta Q_{\gamma\nu}}\right),\quad \sum_i p_i
e_{i\alpha}e_{i\beta} = 0.
\label{eq7}
\end{equation}

Conditions (\ref{eq6})--(\ref{eq7}) are satisfied
by writing the equilibrium distribution functions and
forcing terms as polynomial expansions in the velocity.
The coefficients in the expansion are (in general non-uniquely)
determined by the requirements that these constraints are
fulfilled (see Ref. \cite{lblc} for details). The active contributions
then simply alter the constraints on the second moment of the
$f_i$'s. (Alternatively the derivative of the active term could be
entered as a body force and thus would modify the constraint 
on the first moment of the $p_i$'s; we do not pursue this here.)


In Appendix (\ref{hybridcheckappendix}) we give a quantitative comparison between the hybrid LB algorithm used here and two versions of a fully LB-based code for active nematics \cite{activeLB}. The hybrid code is quite satisfactory in performance; it is also easier to code and runs substantially faster due to the elimination of the cumbersome additional distribution functions required to represent the order parameter dynamics within a fully LB-based approach.

\section{Mapping to Ericksen-Leslie level equations}

In this section, we consider the limit of the equations of motion
(\ref{Qevolution}) and (\ref{navierstokes}) 
when the active molecules are uniaxial, so that
the order parameter can be written in the form
$Q_{\alpha\beta}= q \left
(n_{\alpha}n_{\beta}-\delta_{\alpha\beta}/3\right)$ ($\vec n$ being the
usual nematic director field). We furthermore assume that the magnitude $q$ of the nematic ordering is independent of space and time. The resulting simplified theory is
commonly employed in the physics of active gels (see e.g.
Ref. \cite{kruse1,joanny}); using it, some analytical results have been found.
It is thus useful to explicitly consider this limit (i) to show that
our equations map onto those of Ref. \cite{EPL,joanny} for uniaxial systems,
and (ii) to quantitatively check our numerical results 
against those found analytically for the phase boundaries
separating the active and passive states  \cite{EPL}. In this Section quantities labelled by
``EL'' refer to the resulting director-field model, which is the direct counterpart of the Ericksen-Leslie theory \cite{degennes} of passive liquid crystal hydrodynamics.

\subsection{Order parameter equation of motion}

We first note that the evolution equation (\ref{Qevolution}) of the tensor order
parameter can be written in the usual form
for a purely passive system \begin{equation}
(\partial_t+{\vec u}\cdot{\bf \nabla}){\bf Q}-{\bf S}({\bf W},{\bf
  Q})= \Gamma {\bf H}'
\label{Qevolution_old}
\end{equation}
so long as we write an effective molecular field
\begin{equation}
{\bf H}' = {\bf H} + \frac{\lambda }{\Gamma} {\bf Q}
\label{ha}
\end{equation}
This implies that the the classical linear (in ${\bf Q}$) term
of the molecular field, namely
\begin{equation}
-A_0 \left ( 1-\gamma/3\right) Q_{\alpha\beta},
\end{equation}
is now effectively replaced by
\begin{equation}\label{gamma*}
\left (-A_0 \left ( 1-\gamma/3 \right ) + \frac{\lambda }
{\Gamma}\right )Q_{\alpha\beta}.
\end{equation}
In this manner the ``equilibrium" properties of active nematics
can be said to differ from the passive ones because of the presence of the
active parameter $\lambda$. (This contrasts with the role of $\zeta$, which has no equilibrium counterpart. We will see below, moreover, that the shift created by $\lambda$ has no dynamical consequences in systems where the ordering strength $q$ is fixed.)

After some straightforward algebra (see e.g. \cite{beris} and
Refs. therein), one finds that the linear term now changes
sign for $\gamma=\gamma^*$, with
\begin{equation}
\gamma^* = 3 \left ( 1- \frac{\lambda }{\Gamma
A_0}\right ).
\end{equation}
Similarly the transition point $\gamma=\gamma_c$ for the first-order isotropic-to-nematic transition obeys
\begin{equation}
\gamma_c(\lambda)= \frac{27}{10} \left ( 1 - \frac{\lambda }
{\Gamma A_0}\right) = \gamma_c(0) 
\left ( 1 - \frac{\lambda}{\Gamma A_0}\right).
\end{equation}
Furthermore, for uniaxial nematics with a spatially uniform degree of
ordering $q$ (as assumed at Ericksen-Leslie level -- see above), the solution for $q$
becomes
\begin{equation}\label{order_v_lambda}
q(\lambda)= \frac{1}{4} + \frac{3}{4}\sqrt{1- \frac{8}{3\gamma}+
\frac{8}{3\gamma} \frac{\lambda }{\Gamma A_0}}.
\end{equation}
(Note that this is 3/2 times the largest eigenvalue of the 
${\bf Q}$ tensor.) 

The conventional passive case is recovered by setting $\lambda = 0$ in 
equations (\ref{gamma*}--\ref{order_v_lambda}). 
Note that the value of $q$ at the transition, $q_c = 1/3$, 
is independent of $\lambda$ since it is insensitive to the 
quadratic term of the free energy density. However, the condition for real
solutions (positivity of the term inside the square root) for active nematics is shifted by nonzero $\lambda$ and becomes
\begin{equation}
\bar\gamma= \frac{8}{3} \left ( 1 - \frac{\lambda }{\Gamma A_0}\right).
\end{equation}

\begin{figure}
\begin{center} 
\includegraphics[width=16.cm]{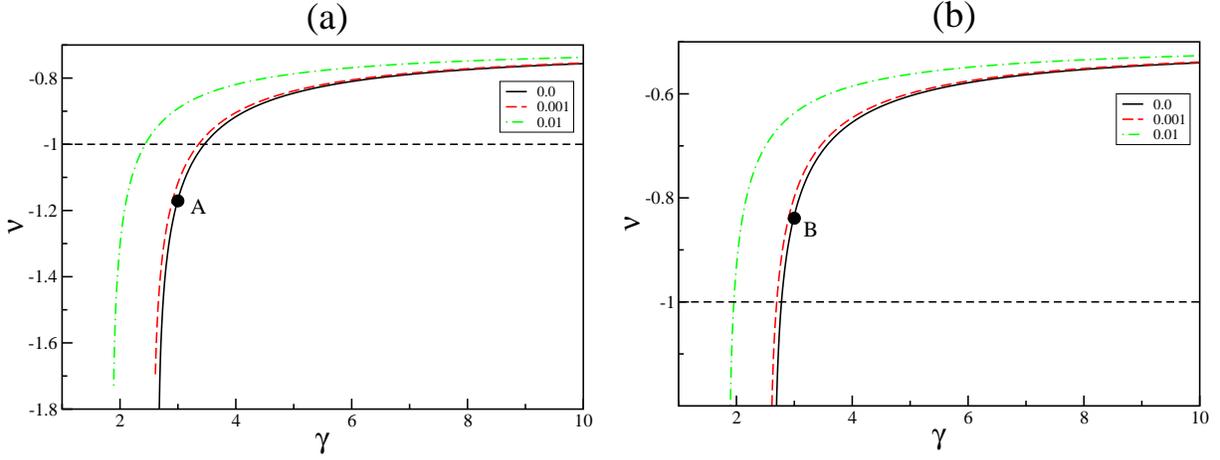} 
\end{center}
\caption{Plot of the $\gamma$ dependence of $\nu_{EL}$;  panels in (a) and (b)
have $\chi = 0.7, 0.5$ respectively. Within each
panel different curves refer to different activity levels $\lambda$ (see legend).  Note that 
for $\xi=0.5$, flow tumbling ($|\nu_{EL}|<1$) is expected throughout the nematic phase. Points A and B represent numerical examples described below.}
\label{fignu_gamma}
\end{figure}

The dynamics of the director field in a uniaxial active liquid crystal of fixed $q$ is controlled by three parameters. These are 
$\gamma_{EL}$, the liquid crystal rotational viscosity;
$\nu_{EL}$, which is another viscosity determining whether the
liquid crystal (in its passive phase) is flow-aligning or flow-tumbling
(for $|\nu_{EL}|$ larger and smaller than 1 respectively);
and $\lambda_{EL}$, which determines the magnitude of 
activity-induced ordering. It is possible to map the dynamical equation of
motion for $Q_{\alpha\beta}$ (\ref{Qevolution_old}) 
onto the model considered
in Ref. \cite{EPL} (the details are worked out in Appendix \ref{workout}), which leads to
the following identifications:
\begin{eqnarray}
\gamma_{EL}&=& \gamma_1=\frac{2 q^2}{\Gamma}\label{gamEPL},\\
\nu_{EL} &=& {\gamma_2\over \gamma_1}=-{(q+2)\xi\over 3q},\label{nuEPL}
\label{gammas_EPL}
\\
\lambda_{EL} &=& 0 .
\end{eqnarray}
These relations show that in our model the dynamics of the tensorial
order parameter may be controlled by tuning $\xi$ and $\Gamma$.
Furthermore, we note that our parameter
$\lambda$ does not control $\lambda_{EL}$ directly, because in Ref.
\cite{EPL} this parameter can already be adsorbed into a Lagrange multiplier introduced to maintain fixed $q$. (To emphasize this, we set it to zero above; see also Appendix \ref{workout}.)  However, changing $\lambda$ in our equations does alter $q$, so qualitatively the meaning of this parameter is similar to that of $\lambda_{EL}$ in \cite{EPL} insofar as it determines the strength of activity-induced self-alignment effects. The relations (\ref{gamEPL}), (\ref{nuEPL}),
give rise to a non-trivial
dependence of the parameter $\nu_{EL}$  on $\gamma$ and $\xi$ as shown
in figure \ref{fignu_gamma}.

\subsection{Navier-Stokes equation}

We now map out the parameters entering the Navier-Stokes
equation (\ref{navierstokes}) onto the analogous equation derived at director-field level in Ref. \cite{EPL}, which is written in
in terms of the ``vectorial'' molecular field $h_\mu$ and of
the director field, $n_\mu$. In 
Ref. \cite{EPL}, the velocity field at steady state of an active gel
is determined by $\nu_{EL}$ (see Sec.\ref{ordersec}), $\eta_{EL}$, which
is an isotropic viscosity similar to the one introduced
in Eq. (\ref{navierstokes}), and $\zeta_{EL}$, which controls the
hydrodynamics in the active phase, determining whether the active liquid
crystal is extensile or contractile as discussed in Section II.
(Note that $\zeta_{EL}$ controls the effect of activity on the Navier-Stokes sector, but does not enter directly the order parameter dynamics as set up in Sec.\ref{ordersec}.)

After some algebra (the details of which are worked out in Appendix B),
we can rewrite Eq.(\ref{navierstokes}) in the required limit of uniaxiality and  fixed $q$. We find that the
six Leslie viscosities for a purely passive liquid crystal
($\lambda=\zeta=0$), which are usually called 
$\alpha_{1,\ldots,6}$ \cite{degennes}, are:
\begin{eqnarray}\label{viscosity1}
\alpha_1&=& -\frac{2}{3\Gamma} q^2(3+4 q-4 q^2)\xi^2,\\
\alpha_2&=& \frac{1}{\Gamma}(-\frac{1}{3}q(2+q)\xi-q^2),\\
\alpha_3&=& \frac{1}{\Gamma}(-\frac{1}{3}q(2+q)\xi+q^2),\\
\alpha_4&=& \frac{4}{9\Gamma}(1-q)^2\xi^2+\eta,\\
\alpha_5&=& \frac{1}{3\Gamma}(q(4-q)\xi^2+q(2+q)\xi),\\
\alpha_6&=& \frac{1}{3\Gamma}(q(4-q)\xi^2-q(2+q)\xi).
\label{viscosity6}
\end{eqnarray}
The Parodi relations,
\begin{eqnarray}
\alpha_3-\alpha_2 &=& {2 q^2\over \Gamma } = \gamma_1,\\
\alpha_6-\alpha_5 &=& -{2\over 3}q\xi({q+2\over \Gamma}) =
\gamma_2,\\
\alpha_2+\alpha_3 &=& \alpha_6-\alpha_5,
\end{eqnarray}
are easily seen to hold.
The Ericksen-Leslie level viscosity and active stress term are
recovered as:
\begin{eqnarray}
\eta_{EL} &=& \eta+\frac{2}{9\Gamma}(q-1)^2\xi^2, \label{etaEPL}\\
\zeta_{EL} &=& \zeta \label{zetaEPL}. 
\end{eqnarray}

\begin{figure}
\begin{center}
\includegraphics[width=10.cm]{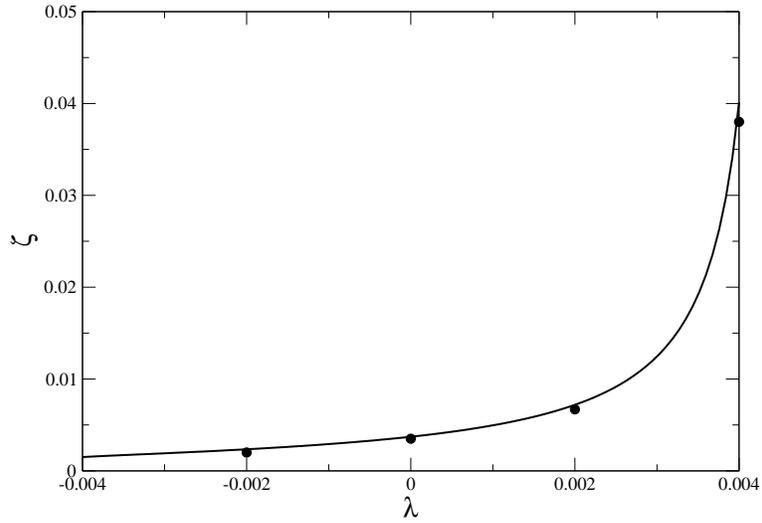}
\end{center}
\caption{ Phase boundary for $L=49$ in the
($\lambda,\zeta$) plane for $\gamma=3.0$, $\tau_f=1$ and $\xi=0.7$.
Four points found numerically from our HLB simulations are also shown (filled circles).} 
\label{fig_phasea}
\end{figure}

\begin{figure}
\begin{center}
\includegraphics[width=10.cm]{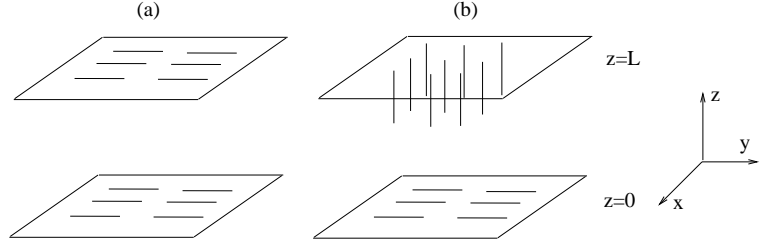}
\end{center}
\caption{ Geometry used for the calculations described in the
text. The active gel is sandwiched between two
infinite plates, parallel to the $xy$ plane, lying at
$z=0$ and $z=L$. We consider (a) normal anchoring and (b) conflicting
anchoring. (The latter would correspond to a hybrid aligned nematic (HAN) cell for a passive liquid
crystal material.)} \label{setup}
\end{figure}

Using the above relations and the results of Ref. \cite{EPL}, 
we obtain the phase boundary in the 
$(\zeta,\lambda)$ plane, for an active nematic confined between parallel plates at separation $L$, with homogenous anchoring at the walls 
(Fig.\ref{setup}): 
\begin{equation}
\zeta L^2 = {{\frac {{12 \pi^2 K} \left(
12\,\tau_f\,\Gamma-5\,
\xi \,{q}^{2}-14\,\xi\,q+ \xi+\xi^2\,{q}^{2}+4\,\xi^2+4
{\xi}^2\,{q}+9 {q}^2 \right) }{ 
9\, \left( \xi\,q+2\,\xi-3\,q \right) }}}
\, \label{Lc}.
\end{equation}
From Eq. (\ref{Lc}) it is apparent that the critical activity
threshold beyond which spontaneous flow is found scales like
$L^{-2}$, and thus vanishes for an infinite system.
Note that the dependence on $\lambda$ of the phase boundary is
indirect, via $q$. Fig. \ref{fig_phasea} shows an example of 
comparison between analytical and simulated phase boundary,
from which it is apparent that there is a good agreement.

\section{Results}

Most of 
the results which we present below refer to a quasi-1D system in which the active nematic is
sandwiched between two plates at separation $L$ in the $z$ direction, with translational invariance assumed in $x$ and $y$ (Fig. \ref{setup}). 
We consider two different boundary conditions:
either homogeneous anchoring along the $y-$direction, or
mixed (conflicting) anchoring at the two plates. We will also refer to
the angle between the director field and the positive $y$ direction
as the polarization angle, $\theta$, the convention being that 
$\theta>0$ if the positive $y$ axis can be superimposed with the director
field with an anti-clockwise rotation of 
an angle $|\theta|$ (which is defined to be smaller than $\pi$), around the
$x$ axis.

\subsection{Spontaneous flow transition in Freedericksz cells}

We first consider homogeneous anchoring where the polarization
at the confining surface is parallel to the $y$-direction,
$\theta=0$. (This geometry is known as the Freedericksz cell in passive
liquid crystal device terminology, \cite{degennes}.) By 
considering Eq. \ref{order_v_lambda} we see that the order
parameter $q$ remains between 0 and 1 for small values of $\lambda$.
Furthermore, we note that for $\xi=0.7$  and $\xi=0.5$ the system is
respectively in the flow-aligning regime (point A in Fig. \ref{fignu_gamma}) and in the flow-tumbling regime (point B). Let us first concentrate on the flow-aligning
regime (point A).
For definiteness we now fix $\lambda=0$, $\tau_f=2.5$, $A_0=0.1$, $K=0.04$,
$\Gamma \sim 0.34$ and $\gamma=3$; while $\xi$ can take on the discrete
values $0.5,0.7$ as just described, and $L$ and $\zeta$ are variable. 
Note that, as described previously, setting $\lambda = 0$ eliminates the 
shift in $q$ arising from self-alignment but this term can anyway be adsorbed into an effective (quasi-passive) free energy. Accordingly, the important activity parameter, for our purposes, is simply $\zeta$.

\subsubsection{Flow-aligning regime}

\begin{figure}
\centerline{\includegraphics[width=10.cm]{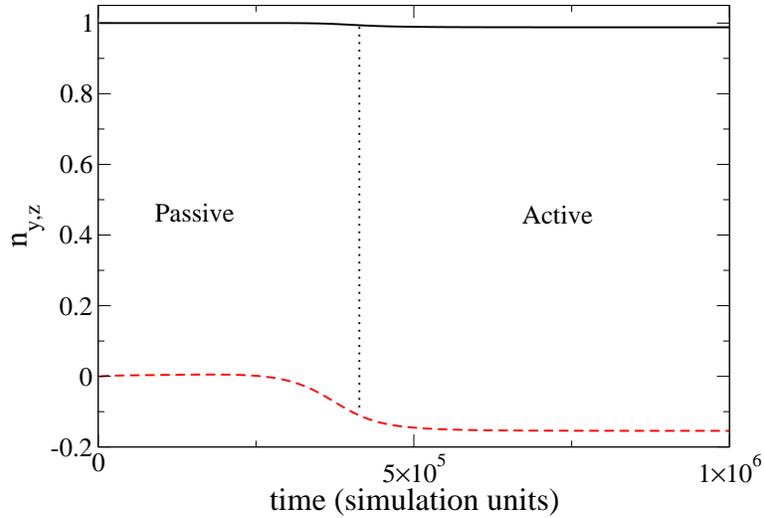}}
\caption{Time evolution of the components of the polarization
field $n_y$ (upper) and $n_z$ (lower), at $z=L/4$. 
Parameters are $L=100$,
$\zeta=0.005$, $\lambda=0$ , $\gamma=3$, $\tau_f=2.5$ and $\xi=0.7$ (flow aligning regime). At the bounding plates, the
field is strongly anchored along the $y$ direction (homogeneous
anchoring). } 
\label{pol_middle_time_9d}
\end{figure}

For  $\xi=0.7$, the system is
flow-aligning and, for instance with $\zeta=0.005$,
the active LC is extensile.  In Figure
\ref{pol_middle_time_9d} we show the time evolution of the
components $n_y,n_z$ of the polarization vector at the center of a
system of size $L=100$ lattice units ($n_x$ is identically zero in this 
case). The polarisation field was 
inizialized along the $y$ direction except for
the midpoint director field, which was initialised with $\theta=10^{\circ}$.  
As one can see for $t>t^*\sim 10^5$ timesteps, the system undergoes a 
transition to an active state, characterised by a spontaneous flow.

\begin{figure}
\centerline{\includegraphics[width=18.cm]{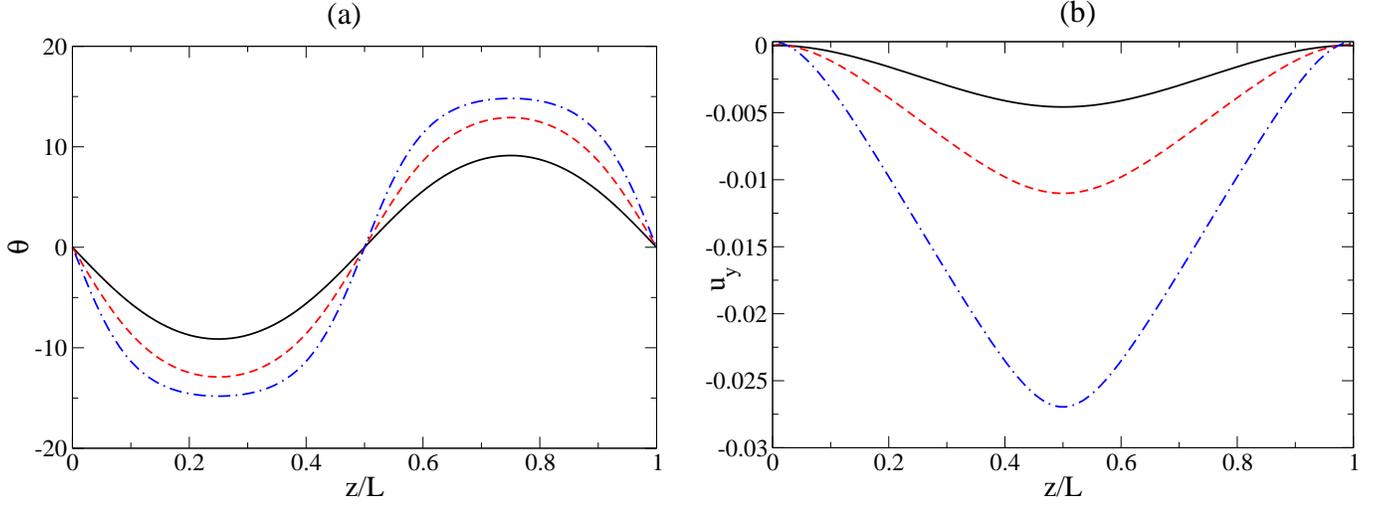}}
\caption{Profiles of director orientation angle (a) and velocity field (b; in lattice units) at steady state
for different values of $\zeta$ in a 
flow-aligning active liquid crystal sample with $L=100$ (other 
parameters as specified in the text). Solid, dashed and dot-dashed
curves correspond to $\zeta = 0.003, 0.005, 0.01$ respectively.
The transition to the active phase occurs at $\zeta=\zeta_c\simeq 0.002$. The flow is bistable: reversing the sign of $\theta$ and $u_y$ together creates an alternative steady-state solution.}
\label{aligning}
\end{figure}

\begin{figure}
\centerline{\includegraphics[width=18.cm]{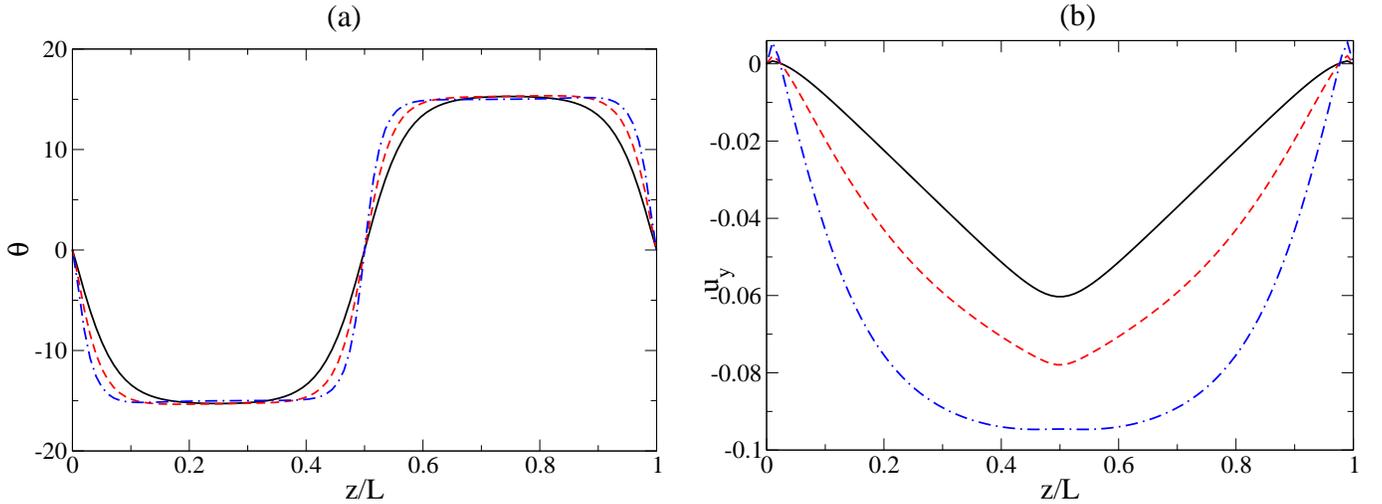}}
\caption{Profiles of director orientation (a) and velocity field (b, lattice units) at steady state
for different values of $\zeta$ in a 
flow-aligning active liquid crystal sample with $L=100$ (other 
parameters as specified in the text). Solid, dashed and dot-dashed
curves correspond to $\zeta = 0.02, 0.04, 0.08$ respectively. All solutions are bistable (see text).}
\label{aligning_deep}
\end{figure}

This happens 
when the scaling variable $\zeta L^2$ becomes larger than the critical 
value found through the solution of Eq. \ref{Lc}.
Thus there are two ways of entering the active phase:
either by increasing the value of $\zeta$ at fixed $L$, or by
increasing the system size at fixed activity.
In Figs. \ref{aligning} and \ref{aligning_deep} we explore the system 
behaviour (respectively director and flow field at steady state)
when the active phase is entered via an increase in the activity parameter $\zeta$.

By means of a stability analysis, valid very close to the 
phase boundary, an analytic expression for $u_y(z)$ was found in
\cite{EPL}. This predicts a sinusoidal modulation with 
a node at the centre of the channel.
While our numerics shows this solution to be metastable for a long time close to the threshold,
the eventual steady state we find is a quasi-Poiseuille flow with a maximum
flow velocity, not a nodal point, at the centre of the channel 
(Fig. \ref{aligning}). Thus with homogeneous boundary conditions and 
assumed translational invariance along the flow direction, we obtain a 
spontaneous net mass flux  rather than the balancing fluxes of forward and 
backward fluid in the two halves of the cell, suggested by the analysis of 
\cite{EPL}. Our numerical simulations thus suggest that the perturbative 
solution is stable at most within a very narrow region close to the phase 
boundary. The overall mass flux is set in a direction chosen by spontaneous 
symmetry breaking or, in practice, small deviations from symmetry between 
$y$ and $-y$ in the initial condition. Note that for a fixed initial condition 
as selected above, the flow direction can also switch on variation in $\zeta$: 
to ease comparisons, some such switches are silently reversed in the figures 
presented here and below. 

\begin{figure}
\centerline{\includegraphics[width=18.cm]{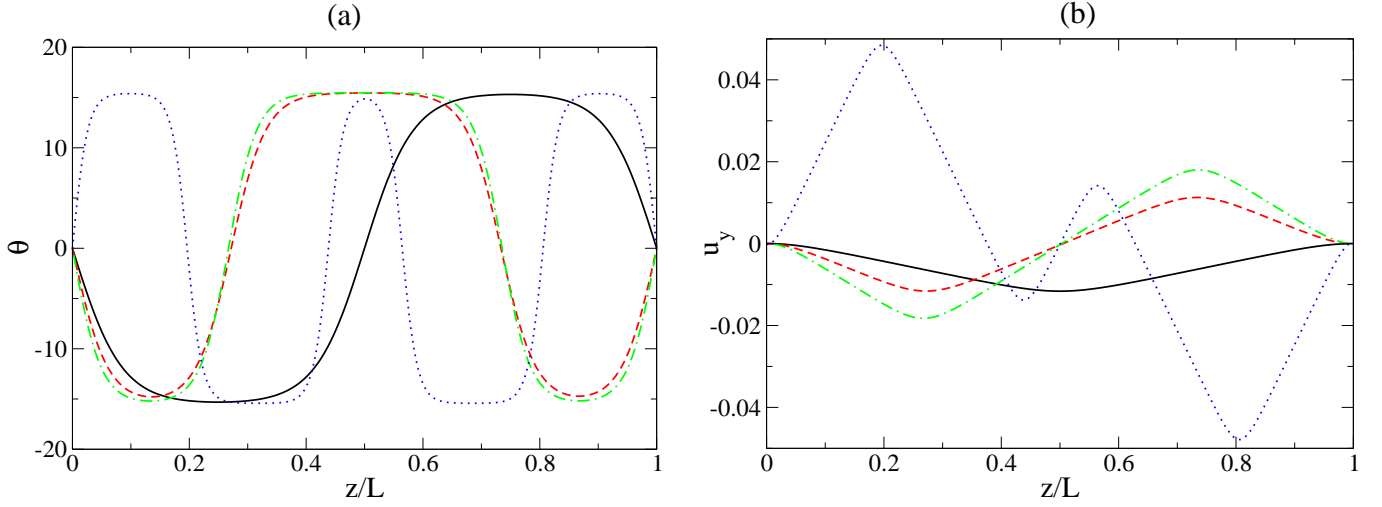}}
\caption{Profiles of director orientation (a) and velocity field (b, lattice units) at steady state
for different values of $\zeta$ in a 
flow-aligning active liquid crystal sample with $L=400$ (other 
parameters as specified in the text). Solid, dashed, dot-dashed
and dotted lines
correspond to $\zeta = 0.001, 0.002, 0.003, 0.01$ respectively.}
\label{aligning400}
\end{figure}

Upon increasing the value of $\zeta L^2$ 
(i.e. moving deeper inside the active phase)
the flow pattern changes from quasi-Poiseuille flow to a ``banded'' flow, with
regions of rather well defined and distinct local shear rates
(Fig. \ref{aligning_deep}). These
{\em bands} (which are clearer and more numerous in larger samples, see Fig. 
\ref{aligning400}) correspond to regions of aligned liquid crystal, which are
separated by sharp interfaces. As the equations deep in the active
phase are strongly non-linear, no analytical results so far exist to
probe the behaviour of an active gel in this regime. The utility of a robust numerical algorithm, as we have developed here with our HLB code, is highly apparent when addressing the potentially complex behaviour in such regimes.
The model we consider allows for a non-constant value of the order
parameter $q$ and we can thus quantify the variations in $q$ that are 
neglected in a director field model. Variations in $q$ are at most
of $1-5\%$ in the simulations reported above, and small dips in the
order parameter correspond to the spatially rapidly varying regions in
the director field profile (i.e. in the ``kinks'' which 
appear at the band edges). Furthermore, these small changes are only 
encountered far from the phase boundary.

\subsubsection{Flow-tumbling regime}

\begin{figure}
\centerline{\includegraphics[width=10.cm]{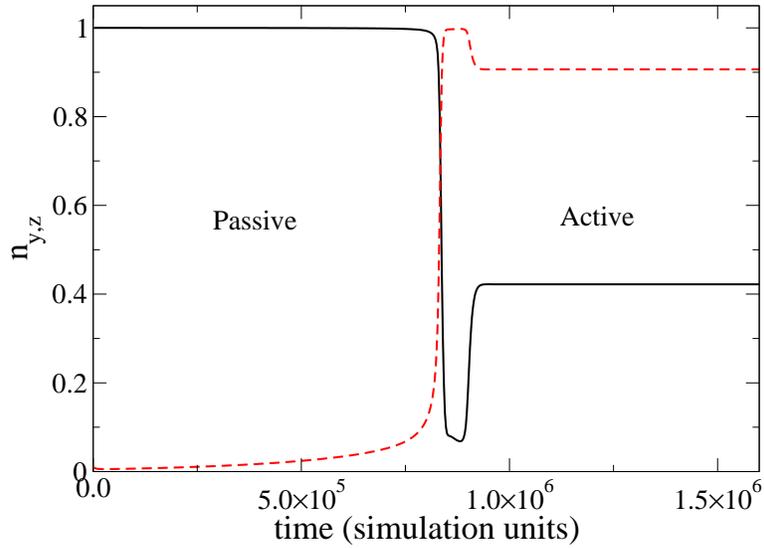}}
\caption{Time evolution of the components of the polarization
field at $z=L/2$ for the flow tumbling case ($\xi = 0.5$). Other parameters are $L=100$, $\zeta=-0.0025$, $\lambda=0$, $A_0 = 0.1$, and $K=0.04$; the transition
as predicted by Eq. \ref{Lc} is at $\zeta=\zeta^*\simeq-0.0022$. 
The director field is
strongly anchored along the $y$ direction (homogeneous anchoring).} \label{pol_middle_time_tumb}
\end{figure}

We now turn our attention to the flow tumbling regime by
considering $\xi=0.5$, $\gamma=3$ and $\lambda=0$.  In this
case Eq. (\ref{Lc}) suggests that, in order to have a
spontaneous flow, $\zeta$ must be negative (i.e. the LC has to
be contractile). This is confirmed by our simulations. 
We consider the value $\zeta=-0.0025$, which is just in the
active phase (see Eq. (\ref{Lc})). In Figure
\ref{pol_middle_time_tumb} we show the time evolution of the
components $n_y,n_z$ of the polarization vector at the center of a
system of size $L=100$, inizialized as for the flow aligning case.
As in the flow-aligning case, for $t>t^*$ the system undergoes a
spontaneous alignment with a consequent spontaneous flow. The time
behavior is however quite different from the one observed in the
flow aligning case. In particular at $t=t^*$ the polarization
vector has an abrupt variation of $\pi/2$ and then reaches a
stationary value  with a polarization angle which strongly
deviates from the starting configuration.

\begin{figure}
\begin{center}
\centerline{\includegraphics[width=18.cm]{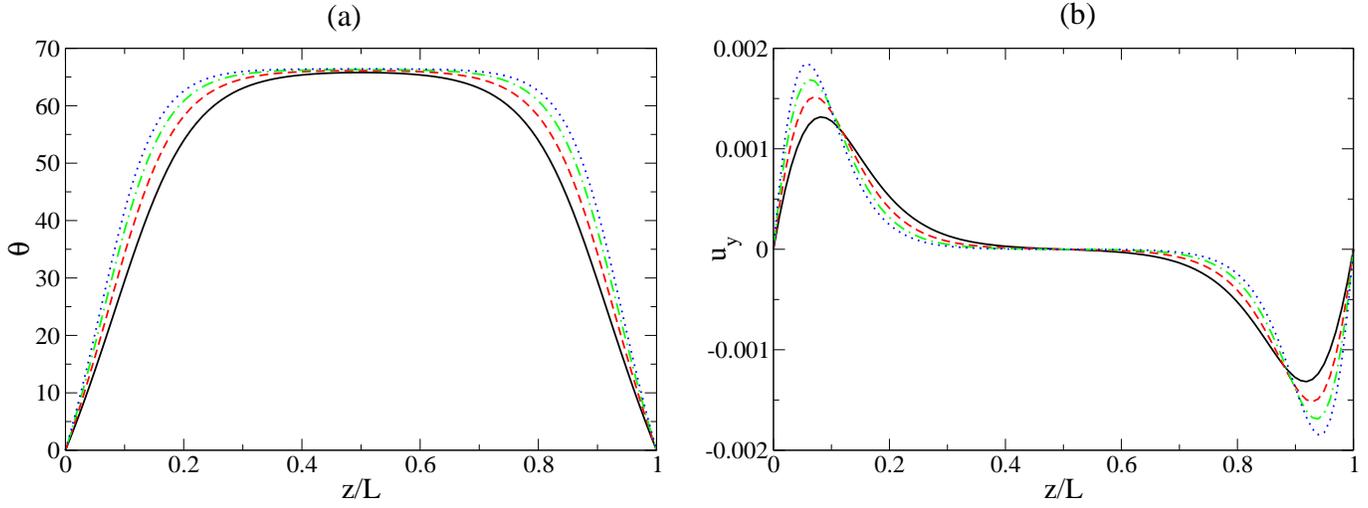}}
\end{center}
\caption{Polarization angle (a) and velocity field (b)
profiles for flow-tumbling active liquid crystals,
with $\zeta$=-0.003 (solid black line), -0.004 (dashed red line),
-0.005 (dot-dashed green line), and -0.006 (dotted blue line).
The transition between the passive and the active phase is
attained at $\zeta=\zeta_c\simeq -0.002$ (see also Eq. \ref{Lc}).} 
\label{tumbling}
\end{figure}

As with the flow-aligning case, we can estimate the critical value
$\zeta_c$ at fixed $L$ (or $L_c$ at a given $\zeta$) 
above which the system starts to display spontaneous flow in
steady state. Again as in the flow-aligning case 
we find good agreement between the value of the
threshold estimated numerically and the analytical prediction
of Eq. (\ref{Lc}).
However, a comparison between the stationary profile of velocity 
and polarization angle profile in the flow-aligning regime and 
in the flow-tumbling one (Figs. \ref{aligning} and \ref{tumbling}
respectively) shows a striking difference. 
While the velocity profile has the shape of
a spontaneous Poiseuille flow for a flow-aligning active liquid
crystal, it is zero in the centre of the channel and confined to the
boundaries in the flow-tumbling case. Also the polarization angle is
quite different: in the flow-aligning case the director
field splays and bends so that the polarization angle approaches
the Leslie values (selected by the local shear), while it is 
almost constant throughout the sample in the flow-tumbling case. 

\begin{figure}
\begin{center}
\centerline{\includegraphics[width=18.cm]{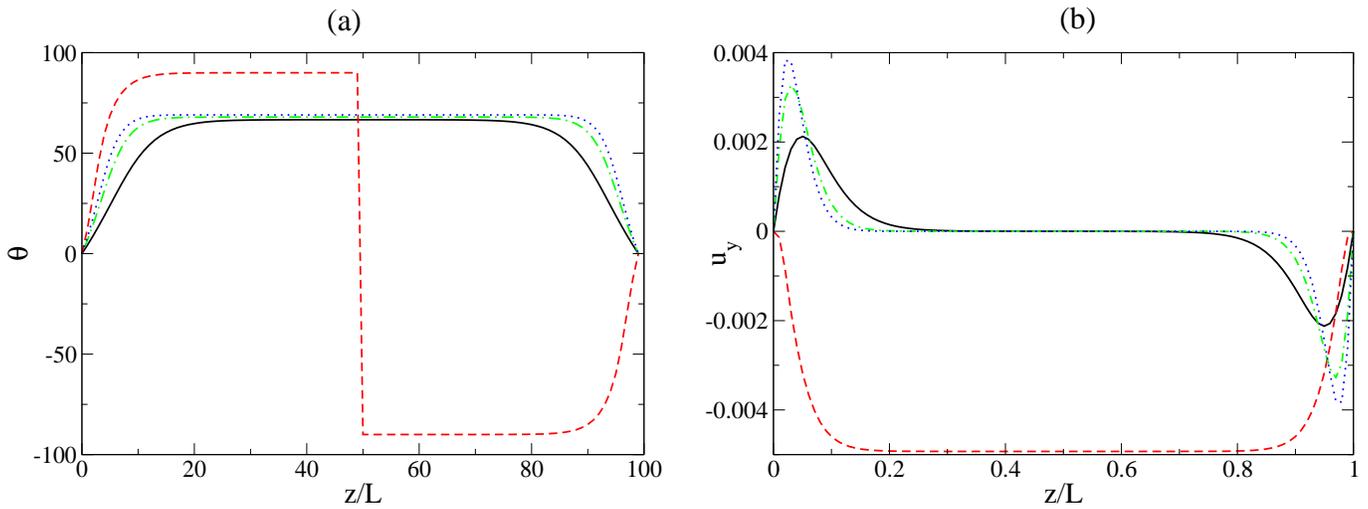}}
\end{center}
\caption{Polarisation angle (a) and 
velocity (b) profiles for flow-tumbling active liquid crystals
deep in the active phase. Curves correspond to
$\zeta$=-0.008 (solid black line), -0.01 (dashed red line),
-0.02 (dot-dashed green line), and -0.03 (dotted blue line).}
\label{tumbling_deep}
\end{figure}

Upon moving deeper inside the active phase, first the velocity field becomes 
confined more and more to the boundaries, while the polarisation angle
becomes increasingly close to $90^{\circ}$ throughout (Fig. 
\ref{tumbling}). For still larger values
of the activity parameter $\zeta$ (Fig. \ref{tumbling_deep}), 
the flow changes sign, passing through an intermediate state 
with plug-like flow in which the polarization has the shape of a kink
(notice however that $\theta=\pm 90^{\circ}$ are equivalent due to
the head-tail symmetry of the director field).
As in the flow-aligning case, order parameter variations are limited 
for $\zeta$ just larger (in absolute value) than the critical value. 
For the simulations presented here and 
deep in the active phase, the order parameter shows some 
drops (similar in magnitude to those found with flow-aligning materials) 
close to the boundary plates, where the shear rates are 
maximal.

\subsubsection{Multi-stability in the active phase}

It is important to consider whether the solutions we have found 
are unique (modulo the trivial bistability associated with sign-reversal, discussed above), or whether each of them is one of many possible
solutions of the equations of motion with given anchoring conditions at the boundary. The selection between such solutions, if they exist, is presumably
governed by the initial conditions. We focus here, for definiteness, 
on the case of contractile active tumbling liquid crystals.

\begin{figure}
\begin{center}
\centerline{\includegraphics[width=18.cm]{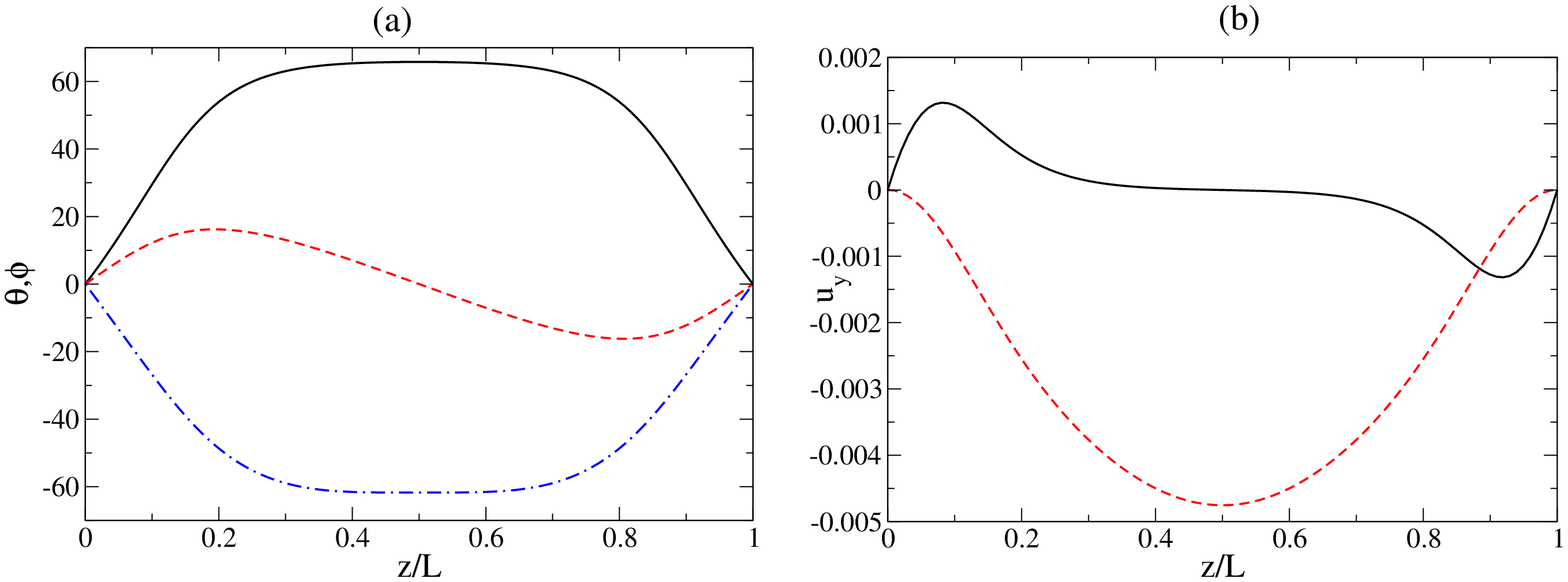}}
\end{center}
\caption{Profiles of director orientation (a) and velocity (b)
for two different steady state solutions found for
contractile tumbling liquid crystals in the active phase 
($\zeta=-0.003$) in the geometry of Fig. \ref{setup}a), starting with two
different initial conditions. In (a) the solid and the dot-dashed
line refer to the two different polarisation angles, while the
long dashed line refers to the $\phi$ angle between the projection of
the director angle onto the $xy$ plane and the positive $x$ axis. 
Initial conditions are given in the text.}
\label{multistable1}
\end{figure}

\begin{figure}
\begin{center}
\centerline{\includegraphics[width=18.cm]{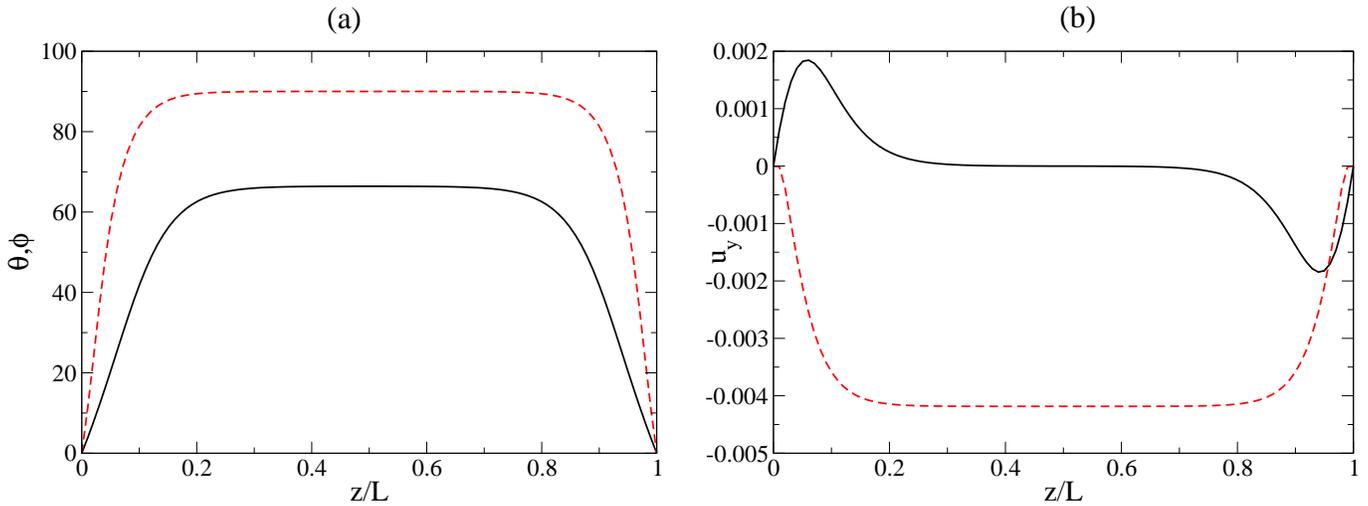}}
\end{center}
\caption{Profiles of director orientation (a) and velocity (b)
for two different steady state solutions found for
contractile tumbling liquid crystals in the active phase 
($\zeta=-0.006$)
in the geometry of Fig. \ref{setup}a). Initial conditions are given in
the text.}
\label{multistable2}
\end{figure}

Figs. \ref{multistable1} and \ref{multistable2} show the results of
two different initial conditions on the steady state 
director and velocity profiles. Fig. \ref{multistable1} shows data for a 
modest value of the activity ($\sim$ 50\% larger in absolute value 
than the critical value to enter the active phase). It can be seen that
one of the solutions has a non-zero component of the director field
along the $x$ direction, so that the director tilts out of the 
``shear plane'' (the $yz$ plane in Fig. \ref{setup}). Fig. \ref{multistable2}
shows another example, deeper in the active phase, in which the
polarisation profiles again differ in steady state for the two different 
initial conditions. One of these initial conditions is the same as
above, for the other we started the director field along the $z$ direction
apart from (the boundary and) the midplane in which the polarisation 
angle was tilted.

Extensile aligning liquid crystals behave in a similar way. As a rule of 
thumb, multistability appears to increase for {\em intermediate} values of the
activity. For the cases considered here, we only find a single (bistable) solution in the 
active phase close to the phase boundary and again for very large activity.
It should be noted that also passive liquid crystals can have metastable
multiple solution in equilibrium (for instance super-twisted structure
are metastable). However, in that case (in the presence of thermal noise, and in the absence of external driving) one can speak of a ``most stable solution'' which is
unambiguously determined by free energy minimization. No such criterion exists for our non-equilibrium problem,
as the equations of motion 
cannot be written down completely in terms of a free energy. (Note however that, were $\zeta = 0$, this could be done even in the presence of the active self-alignment term $\lambda$.)
\if{
We note finally that multistability starting from different boundary
conditions is expected to lead to chaotic dynamics, at least away from
the phase boundary in the active phase.
}\fi

\subsection{Spontaneous flow in hybrid aligned nematic cells}

Now we consider a hybrid-aligned nematic cell (HAN cell, in passive liquid
crystal terminology \cite{rheoHAN}), 
in which the polarization vector is anchored
homogeneously at $z=0$ and homeotropically at $z=L$.
We restrict attention to $\xi=0.7$, the flow-aligning case.

\begin{figure}
\begin{center}
\centerline{\includegraphics[width=18.cm]{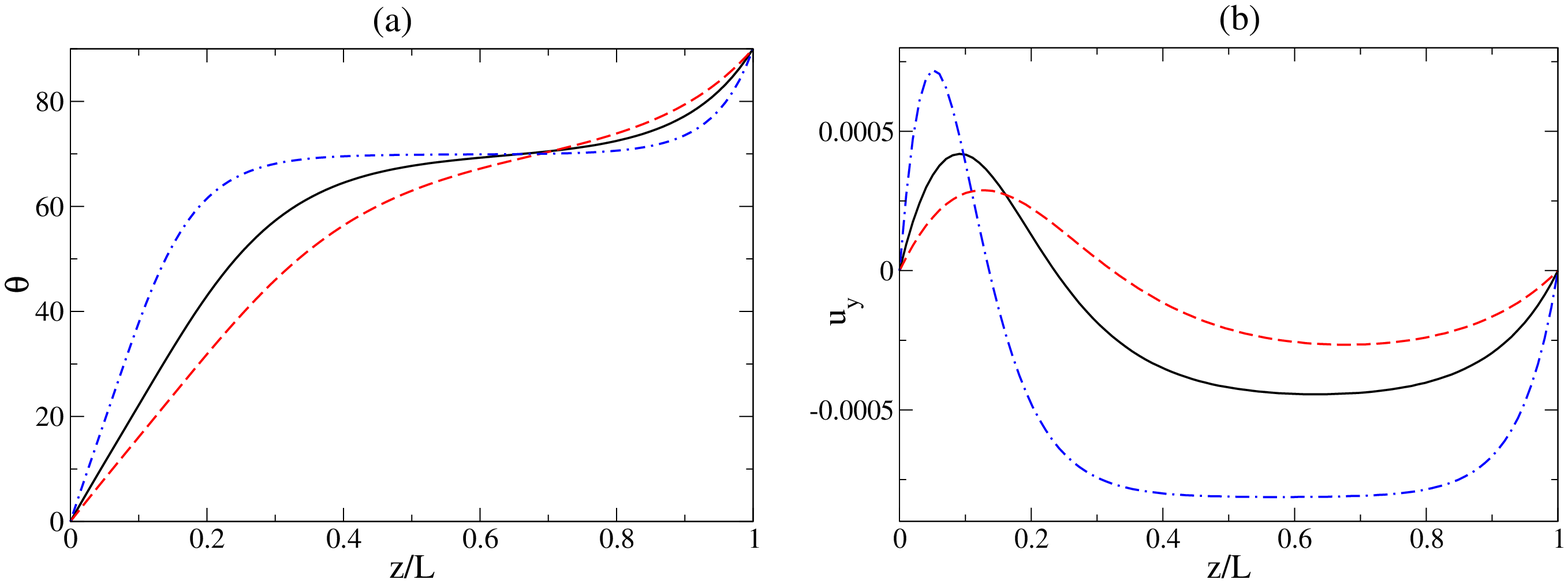}}
\end{center}
\caption{Profiles of director orientation (a) and velocity (b)
for flow-aligning contractile active liquid crystals
in a HAN geometry. Curves correspond to
$\zeta=-0.001$ (solid black line), $-0.0005$ (dashed red line),
$-0.003$ (dot-dashed blue line).}
\label{HANcontractile}
\end{figure}

Unlike the Freedericksz cell, the conflicting anchoring now 
leads to an elastic distortion in equilibrium
even within the passive phase of the active system (as it would in a strictly passive nematic). As a result {\em any} non-zero value
of $\zeta$, whether positive or negative, leads to spontaneous
flow in steady state, as the active pressure tensor is no
longer divergence-free when $\zeta\ne 0$. Thus even
contractile aligning liquid crystals flow spontaneously
in this geometry (Fig. \ref{HANcontractile}). The velocity profiles
in steady state in this case show extended regions with very low shear rate
and plug-like flow, coexisting with strongly sheared ``boundary layers''.
This is similar to what was observed in Section IV A.1 for 
contractile (tumbling) liquid crystals in a Freedericksz cell geometry.
The region of the cell in which the director field is close to
homeotropic anchoring ($\theta=0$) increases with $|\zeta|$. 

\begin{figure}
\begin{center}
\centerline{\includegraphics[width=18.cm]{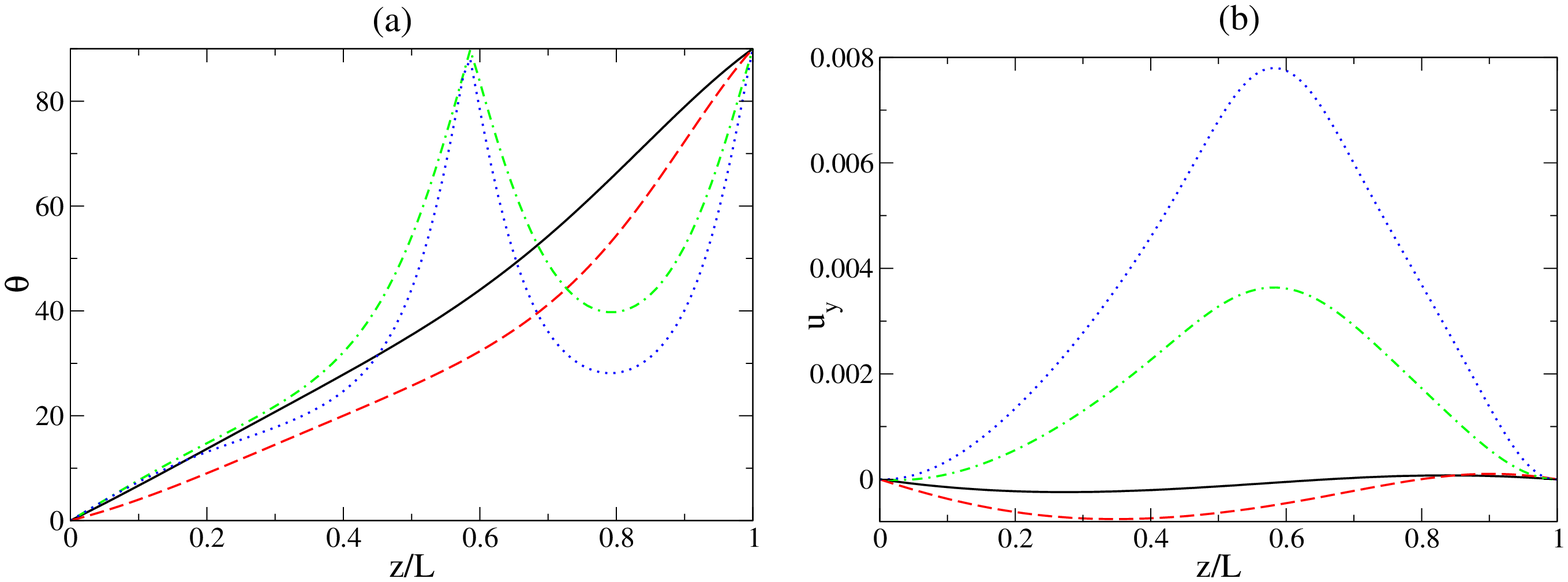}}
\end{center}
\caption{Profiles of director orientation (a) and velocity (b)
for flow-aligning extensile active liquid crystals
in a HAN geometry. Curves correspond to
$\zeta=-0.001$ (solid black line), $-0.0005$ (dashed red line),
$-0.003$ (dot-dashed blue line).}
\label{HANextensile}
\end{figure}

\begin{figure}
\begin{center}
\centerline{\includegraphics[width=18.cm]{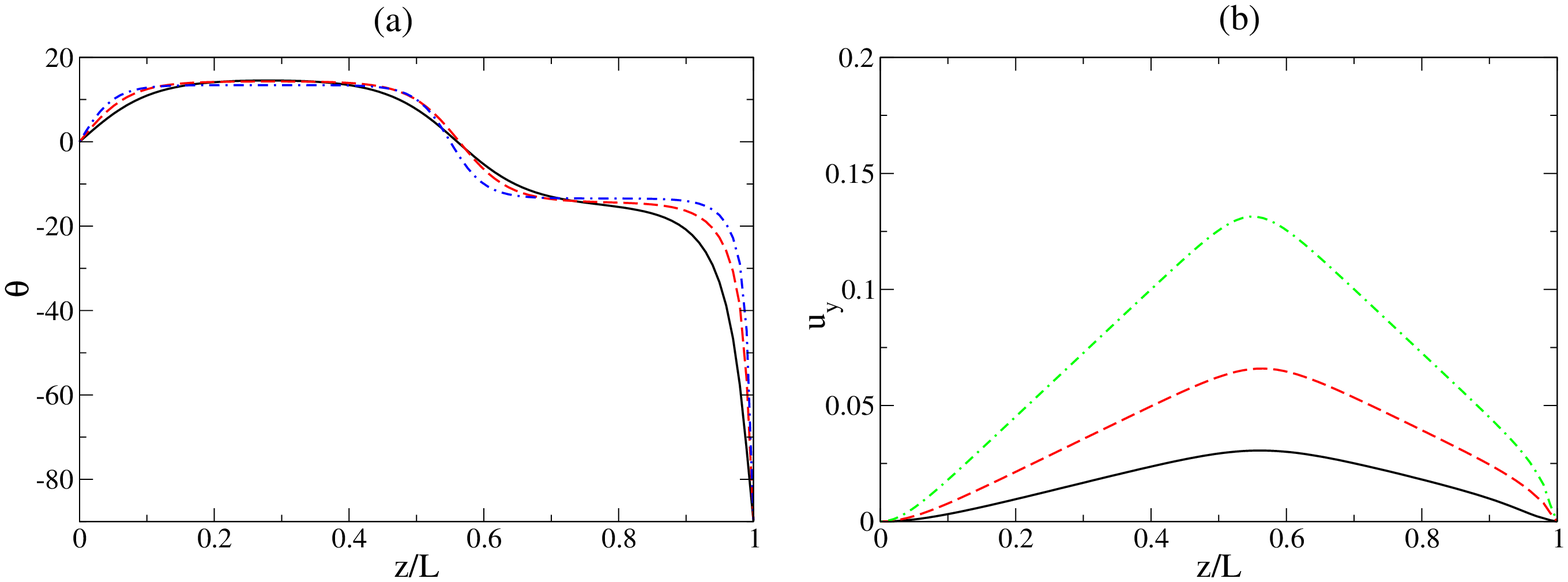}}
\end{center}
\caption{Profiles for director orientation (a) and velocity (b)
for flow-aligning extensile active liquid crystals
in a HAN geometry. Curves correspond to
$\zeta=-0.001$ (solid black line), $-0.0005$ (dashed red line),
$-0.003$ (dot-dashed blue line).}
\label{HANextensile_deep}
\end{figure}

The behaviour of extensile aligning materials in a HAN geometry is reported
in Figs. \ref{HANextensile} and \ref{HANextensile_deep} for smaller and 
larger values of $\zeta$ respectively. The spontaneous
flow is asymmetric. Initially there are oppositely flowing slabs
of liquid crystals, which distort the director field by
creating homogenously aligned region separated by thin regions
of homeotropic ordering. These profiles are then
supplanted by an asymmetric quasi-Poiseuille flow, which  
resembles the response of a purely passive HAN cell to a pressure difference 
driven flow \cite{rheoHAN}. At larger values of $\zeta$
the director profile throughout is close to the one obtained
for a Freedericksz cell, with only a highly distorted 
boundary layer to satisfy the homeotropic anchoring at the
top plane ($z=L$).

\subsection{Spontaneous flow in two dimensions}

Thus far, all simulations reported here were performed
in a quasi-1D geometry, where translational invariance is assumed along $x$ and $y$.  The same simplification is often employed in numerical studies of passive liquid
crystals (see many examples in Ref. \cite{degennes},
as well as e.g. Refs. \cite{ramaswamy_chaos,rheoHAN} for rheological
studies); 
moreoever, as shown above they allowed us to check detailed analytical predictions (calculated at director-field or EL level) in exactly this geometry \cite{EPL}.
It is clearly important and interesting to consider whether there are additional 
spontaneous flow instabilities in a higher dimensionality. With periodic boundary conditions such instabilities must spontaneously break the translational invariance in $x$ and $y$; we limit our attention to this case, but note that confining cell walls might also play an important role.

We next present 2D simulations ($L_z = 100, L_y = 100, L_x = 0$) in which we again have two parallel plates, normal to $z$; 
translational invariance along $x$ is maintained but periodic boundary conditions are used to allow breakdown of this along the flow direction, $y$.
We initialised the simulation with the director field along the 
$y$ direction except for points along the mid-plane $z=L/2$, in which
there was an alternating tilt of $\pm 10^{\circ}$ in stripes (the width of the
initial stripes did not affect the steady state reached at the end of the
simulations).

\begin{figure}
\begin{center}
\centerline{\includegraphics[width=16.cm]{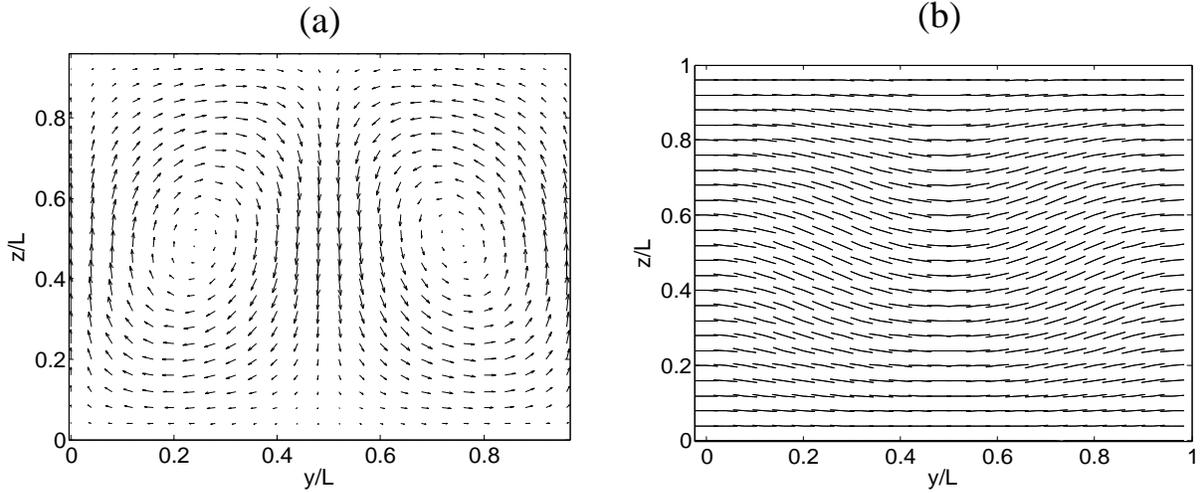}}
\end{center}
\caption{Maps of velocity field (a) and director field (b) in steady state for
an active aligning liquid crystal with $\zeta=0.001$ (extensile),
simulated on a two-dimensional $L=100 \times L=100$ grid.}
\label{2d_flow1}
\end{figure}

Fig. \ref{2d_flow1} shows results for a moderate value
of the activity parameter $\zeta$ (0.001), for which the liquid crystal
enters the spontaneously flowing active phase. Spontaneous flow appears as
a pair of convection rolls which lead to a splay-bend in-plane deformation
of the director field profile. The order parameter is to a good approximation
constant ($q \simeq 0.5$) throughout the sample. The threshold at which the spontaneous flow appears is smaller than the
one found in the quasi-1D simulation (for which with the same parameters
$\zeta_c\simeq 0.002$, see above). This is due to the fact that along 
$y$ effectively homeotropic anchoring conditions are seen, and the active
phase is entered for a smaller value of $\zeta$ in this geometry.
Note that, since at onset of the convection rolls there are exactly two of these in the periodic cell, the details of the transition may now depend sensitively on the aspect ratio of the cell.

\begin{figure}
\begin{center}
\centerline{\includegraphics[width=15.cm]{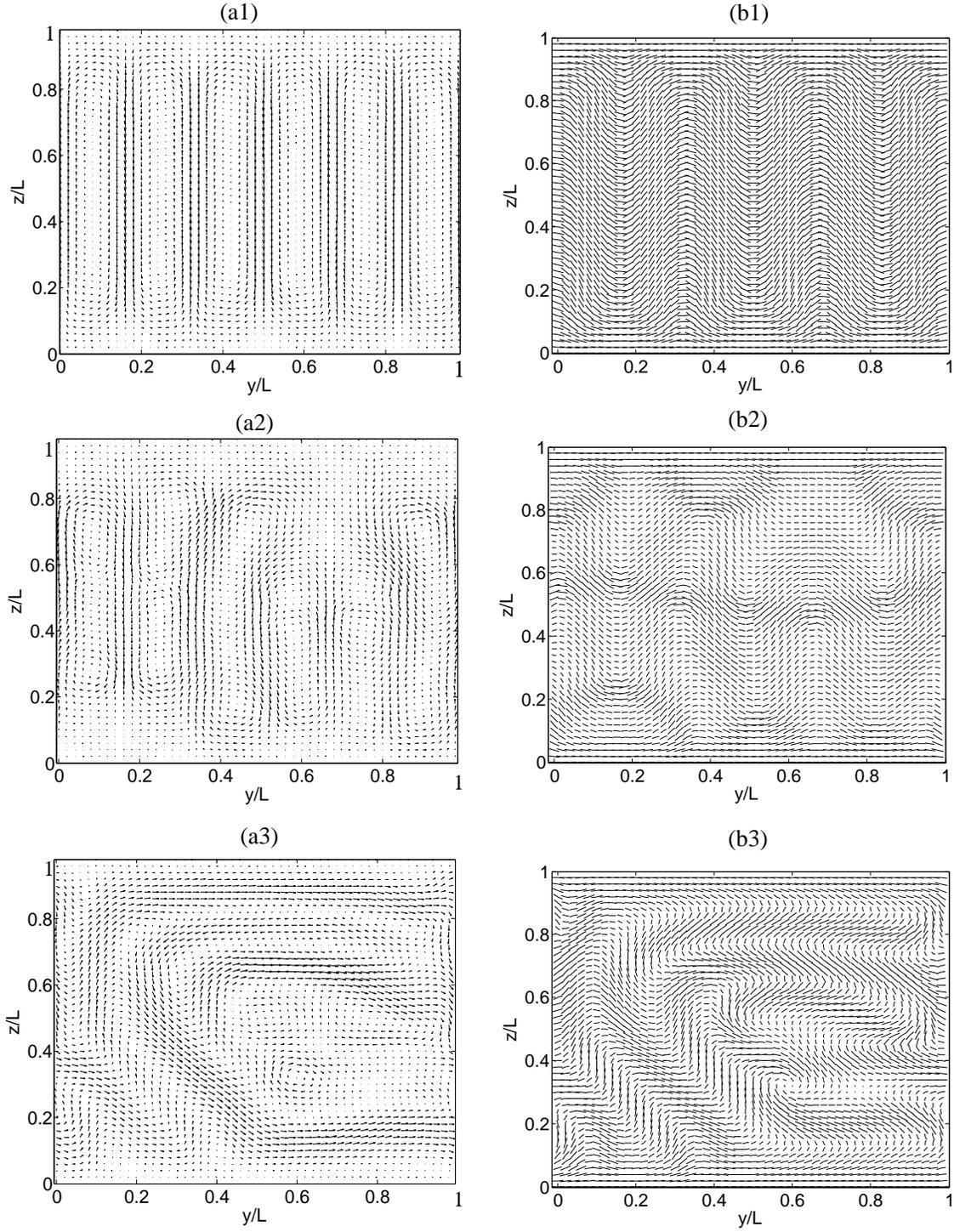}}
\end{center}
\caption{Maps of velocity field (a1-a3) and director field (b1-b3) for
an active aligning liquid crystal with $\zeta=0.01$ (extensile), simulated
on a two-dimensional $L=100 \times L=100$ grid. The three
rows correspond to the configurations after $10^4$,
$3\times 10^4$, $10^5$ lattice Boltzmann steps respectively.}
\label{2d_flow2}
\end{figure}

As we go deeper into the active phase, the number of convection rolls
is, at early times in the simulations, larger 
(Fig. \ref{2d_flow2} (a1,b1)). These convection rolls then split up,
and the flow field acquires an out-of-plane component (i.e. there is flow along the
$x$ direction). After this happens, a number of vortices form which lead to
a complicated flow which is accompanied by the formation of defects
(of topological strength $\pm 1/2$) in the director field
profile. The simulation, followed in Fig. \ref{2d_flow2}, 
does not lead to a steady state. It would seem plausible that
the corresponding trajectories in phase space may be chaotic, but we have not attempted to test this directly. Moreover, once a nonzero $x$ velocity has been acquired, there is a strong possibility of breakdown of translational invariance in $x$; to explore this would require fully 3D simulations.
Note however that in this regime the structural length scale of the flow appears small on the scale of the simulation cell and therefore might cease to be sensitive to its shape.

\section{Discussion and conclusions}

We have presented a hybrid lattice Boltzmann algorithm to solve the 
equations of motion of an active nematic liquid crystal.
In our equations the orientational degrees of freedom 
are characterised by a tensorial order parameter. This renders
our algorithm general enough to deal -- in principle -- with non-homogeneous,
flow-induced or paranematic ordering, as well as with topological
defects. The model we analyse is equivalent to the one proposed in Ref. 
\cite{ramaswamy}.

Our main results are the following. First, we have explicitly
mapped our model onto the one considered in Ref. \cite{EPL} in the
limiting case of a uniaxial liquid crystal with a spatially uniform 
and time independent magnitude of ordering. This is useful when comparing 
the different approaches which are now being proposed to study the
physics of active materials.

Second, we found a spontaneously flowing phase (active phase) for
a wide range of values for the activity parameter $\zeta$ in a quasi-1D geometry
where the director field is constrained to lie along a common direction along both confining plates. (A second activity parameter, $\lambda$, merely renormalizes the equilibrium parameters of the passive material.) Our simulations confirm the location of the phase transition from passive to active phase found via a linear stability analysis in Ref. \cite{EPL}, but show that, for a wide range of parameters within the active phase, even very close to the boundary, the spontaneous flow profile has a quite different symmetry from the one predicted by that analysis. Instead of a sinusoidal flow with a node at the midplane, flow-aligning and flow-tumbling liquid crystals display 
a quasi-Poiseuille flow and a ``boundary layer'' type flow respectively. (Both flow profiles are bistable.) 

Our numerical method can readily probe, for the first time, the hydrodynamic behaviour of active materials
deep in the active phase, where we gave evidence of a spontaneously banded flow
for the flow-aligning case. Far from the phase boundary, there are multiple
(initial condition dependent) solutions, and the system displays hysteresis.

Third, if conflicting (HAN-type) anchoring conditions are applied at the confining plates,
spontaneous flow occurs for any values of the activity parameter $\zeta$, however small. 
Finally, we performed two-dimensional simulations, with periodic boundary conditions along
the $y$ direction and planar anchoring along that direction on both confining plates. These suggest that there are additional instabilities
in a quasi-2D geometry. Moreover, at high activity levels, there can 
also be a spontanous flow also in the $x$ direction in this geometry.

These results demonstrate a remarkable richness in the steady-state hydrodynamic behaviour of active nematic materials, even in the absence of exernal drive such as an imposed shear flow. (As such, they have no counterpart in the physics of passive nematics.) Our hybrid lattice Boltzmann methodology, which combines LB for momentum with finite difference methods for the order parameter tensor $Q_{\alpha\beta}$, offers a robust and efficient method for probing these effects. It can equally well handle transient phenomena, some of which we explored above, and can readily be modified to allow for imposed flow.

Our algorithm can be generalized in several ways. For instance, an additional
order parameter equation, describing the time evolution of a polar vector
field, can be considered with little more effort. This would allow a 
full 2D study of polar active nematics \cite{liverpool} 
with a variable degree of ordering.
Similarly, chiral active liquid crystals
can be straightforwardly treated \cite{active_chiral_gel}, 
for instance to model concentrated actomyosin
solutions. 
Actin fibers in
very concentrated solutions undergo a nematic to cholesteric transition; another
candidate for an active chiral liquid crystal
might be a solution of DNA fragments interacting with polymerases or
other motors \cite{peter}. Also, it would be of interest to
use the present algorithm to characterise the rheological properties
and map out the flow curves of an active liquid crystal under imposed shear.
We shall report on such work in future publications. We also hope to report soon on fully three-dimensional simulations of active materials, along the lines pioneered for passive nematics in \cite{lblc}.

We acknowledge EPSRC for support, and are grateful to L. Tubiana for
a critical reading of this work.

\appendix
\section{Comparison of hybrid with conventional LB codes}
\label{hybridcheckappendix}
\begin{figure}
\begin{center}
\includegraphics[width=16.cm]{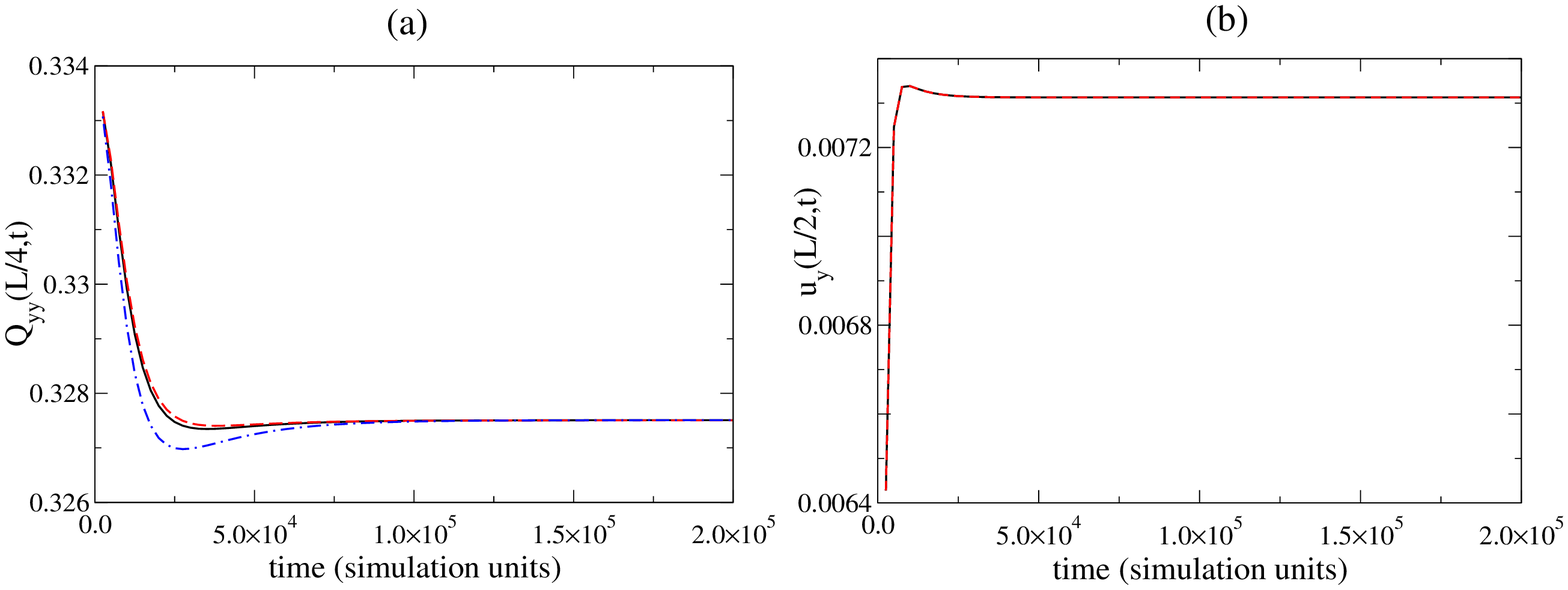}
\end{center}
\caption{Time evolution of $Q_{yy}$ at $z=L/4$ (a) and of $u_y$ in the
mid-plane (b), as predicted by our hybrid LB treatment (solid black lines),
and by two types of full LB treatment (dashed red lines, with
the double gradient terms entered in the first moment constraint,
to avoid spurious velocities at equilibrium; and
dot-dashed blue lines, with the double gradient term entered in the
second moment constraint).}
\label{hybrid_check}
\end{figure}

In Fig. \ref{hybrid_check} we show the director and velocity dynamics 
at $z=L/4$ (in the geometry of Fig. \ref{setup}) and in the mid-plane 
respectively,  computed via the hybrid algorithm discussed in this paper 
and via a full LB algorithm (as described in Refs. \cite{colin,lblc} for passive nematics and in \cite{activeLB} for the active case).
The agreement proves the validity of our hybrid approach. 
Note that two full LB algorithms are benchmarked against the
hybrid code. In one case the double gradient term is entered as
a constraint in the second moment, in the other its derivative is
entered as a body force (this second procedure guarantees that no
spurious velocities are found in steady state, see e.g. Ref. \cite{nidhal}).
It can be seen that the LB treatment with the double gradient terms 
entered in the second moment constraint leads to a small deviation at
intermediate times. This we interpret as a discretisation error, as this
method in 2D is known (for conventional i.e. passive liquid crystals) to lead to discretization errors causing small spurious
velocities even in the steady state \cite{nidhal}.

\section{``Ericksen-Leslie'' limit of the order parameter evolution equation}
\label{workout}

In this Appendix we map the order parameter evolution equation
used in this work, Eq. \ref{Qevolution}, onto the analogous
equation used in Ref. \cite{EPL}, by taking the limit of a uniaxial
liquid crystal with spatially uniform and temporally constant
magnitude of ordering $q$. In this way we will recover Eqs. 
\ref{gamEPL} and \ref{nuEPL}.

To this end let us first write the  ${\bf Q}$ evolution equation
(\ref{Qevolution}) for ${\bf H}$. This gives, formally,
\begin{equation}
\Gamma{\bf H} =
(\partial_t+{\bf u}\cdot{\bf \nabla}){\bf Q}-{\bf S}({\bf W},{\bf
  Q})-\lambda{\bf Q}.
\label{Qevolution1}
\end{equation}
By considering the uniaxial expression for ${\bf Q}$ (see
Eq. \ref{uniaxial}) we obtain
\begin{eqnarray}
\Gamma H_{\beta\mu}&=&  (\partial_t q)n_\beta n_\mu
-\frac{\delta_{\beta\mu}}{3}\partial_t q +
\left(u_{\gamma}\partial_{\gamma}q \right)n_\beta n_\mu
-\frac{\delta_{\beta\mu}}{3}\left(u_{\gamma}\partial_{\gamma}q \right)
\delta_{\beta\mu}
\nonumber\\
&&
+q\left(\partial_t n_{\beta}\right)n_{\mu} +
q n_{\beta}\left(\partial_t n_{\mu}\right)+
q\left(u_{\gamma}\partial_{\gamma}n_{\beta} \right)n_\mu+
qn_{\beta}\left(u_{\gamma}\partial_{\gamma}n_{\mu} \right)
 \nonumber\\
&&
-\lambda q n_{\beta}n_{\mu} + \lambda q \frac{\delta_{\beta\mu}}{3} + \frac{2}{3}\xi(q-1)D_{\beta\mu}\nonumber\\
&&
-\xi q\left (D_{\beta\gamma}n_{\gamma}n_{\mu}+n_{\beta}n_{\gamma}D_{\gamma\mu} \right )
-q\left(\Omega_{\beta\gamma}n_{\gamma}n_{\mu}-n_{\beta}n_{\gamma}\Omega_{\gamma\mu} \right)\nonumber\\
&&
+2q\xi n_{\beta}n_{\mu}Tr({\bf{Q}\bf{W}})-\frac{2}{3}\xi (q-1)Tr({\bf{Q}\bf{W}}).
\label{H1}
\end{eqnarray}
As can be easily checked, one can substitute ${\bf W}$
with ${\bf D}$ in (\ref{H1}). As we have assumed that $q$ does
not depend on $t$ and ${\vec{r}}$, we obtain:
\begin{eqnarray}
\Gamma H_{\beta\mu}&=& q (n_\mu
  N_\beta+n_\beta N_\mu)-q \xi(D_{\beta\gamma}n_\gamma
  n_\mu+n_\beta n_\gamma D_{\gamma\mu})
-\lambda q\left(n_\beta n_\mu -\frac{\delta_{\beta\mu}}{3}\right)\nonumber\\
& & \quad  +{2 \over 3}(q-1)\xi
  D_{\beta\mu}+{1 \over 2} q^2 \xi n_\beta n_\mu
  D_{\gamma\nu}n_\nu n_\gamma+{2 \over 3} q(1-q)\xi
  \delta_{\beta\mu}D_{\gamma\nu}n_\nu n_\gamma
\label{H2}
\end{eqnarray}
where $N_\beta, N_\mu$ are co-rotational derivatives defined as,
\begin{eqnarray}
N_\beta
&=& \partial_t n_\beta+u_\gamma \partial_\gamma n_\beta+\Omega_{\beta\gamma}n_\gamma \nonumber\\
&=&\partial_t n_\beta+u_\gamma \partial_\gamma n_\beta-({\bf
  \omega} \times {\bf n})_\beta
\end{eqnarray}
and ${\bf \omega}=\nabla \times {\bf u}/2$. In order to write the
evolution  equation (\ref{H2}) in a form that resembles the one
introduced in \cite{EPL} we note first that, by the chain rule,
\begin{eqnarray}
h_\mu&=&-{\delta {\cal F} \over \delta n_\mu}= -{\delta {\cal F}
\over \delta Q_{\alpha\beta}} {\partial
  Q_{\alpha\beta}\over \partial n_\mu} \nonumber\\
&=& H_{\alpha\beta}q(n_\beta \delta_{\alpha\mu}+ n_\alpha
\delta_{\beta\mu}) \nonumber\\
&=& q(n_\beta H_{\beta\mu}+ n_\alpha H_{\alpha\mu})=2q(n_\beta
H_{\beta\mu}) \label{h(n,H)}.
\end{eqnarray}
If we now multiply (on the left) both members of Eq. (\ref{H2}) by
$n_\beta$ and we use the constraint $n_\beta n_\beta=1$ we obtain
after some algebra,
\begin{equation}
\Gamma h_\mu/2q=q N_\mu-{1\over 3}(q+2)\xi n_\gamma D_{\gamma\mu}
- \frac{2}{3}\lambda q n_\mu
\label{eqa1}
\end{equation}
where we have omitted terms $O(n^3)$. Clearly, if
$\lambda=0$, Eq. (\ref{eqa1}) reduces to the usual Ericksen-Leslie
equation for the director field, namely \cite{degennes}
\begin{equation}
h_\mu = \gamma_1 N_\mu+\gamma_2 n_\alpha D_{\alpha\mu}
\label{le1}
\end{equation}
where
\begin{eqnarray}
\gamma_1 &=& \frac{2 q^2}{\Gamma},\\
\gamma_2 &=& -{2q\over 3\Gamma}(q+2)\xi. \label{gammas}
\end{eqnarray}
If, on the other hand, the active term $\lambda \ne 0$ we have
\begin{equation}
N_\mu=\frac{\Gamma}{2q^2} h_\mu +{1\over 3}\frac{(q+2)}{q}\xi
n_\gamma D_{\gamma\mu}.
\label{eqa2}
\end{equation}
Note that terms proportional to $n_\mu$ drop out of the equations
in this mapping. Indeed they contribute a component of the molecular field
parallel to $n_\mu$, which would tend to increase the magnitude of
the director field $q$. This is prevented by the Lagrange multiplier
which appears in the vectorial ``Ericksen-Leslie'' model (to maintain 
constant $q$). As a result such terms simply
change the relationship between the Lagrange multiplier and the
magnitude of order and not the structure of the director
field equation.
By comparing Eq. (\ref{eqa2}) with  Eq. (3) of \cite{EPL} (there
$Dn_\mu/Dt=N_\mu$) we then obtain the relations listed in
Eqs. (\ref{gamEPL}) and (\ref{nuEPL}) in the text.

\section{``Ericksen-Leslie'' limit of the Navier-Stokes equation}

In this Appendix we work out the details of the mapping
between the Navier-Stokes equation in our tensorial model 
in the uniaxial limit of constant $q$, and the momentum
balance equation used in the ``Ericksen-Leslie'' version of
Ref. \cite{EPL}, which was reported in Section III B in the text.
To this end, we need to write the total stress tensor
$\Pi_{\alpha\beta}=\Pi^{passive}_{\alpha\beta}+\Pi^{active}_{\alpha\beta}$
in terms of the molecular and director fields, $h_\mu$ and $n_\mu$
respectively, which are used in director field based models.
As in Appendix B we write $Q_{\alpha\beta}$ in uniaxial
form i.e ${\bf Q}=q({\bf P}-{\bf I}/3)$ where
$P_{\alpha\beta}=n_\alpha n_\beta$. Note that ${\bf P}^2={\bf P}$
and $Tr ({\bf P})=1$ and recall that the Ericksen-Leslie
expression for the total stress is:
\begin{eqnarray}
\sigma_{\alpha\beta}^{EL} &=& \alpha_1 n_\alpha n_\beta n_\mu n_\rho
D_{\mu\rho} + \alpha_4 D_{\alpha\beta}+\alpha_5 n_\beta n_\mu
D_{\mu\alpha}
\\ \nonumber
& & \quad+\alpha_6 n_\alpha n_\mu D_{\mu\beta}+\alpha_2 n_\beta
N_\alpha+\alpha_3 n_\alpha N_\beta \label{ELstress}.
\end{eqnarray}
We first consider the anti-symmetric part of the passive stress tensor
in the tensorial model, namely:
\begin{eqnarray}
\tau_{\alpha\beta}&=&{\bf Q}\cdot{\bf H}-{\bf H}\cdot{\bf Q}\nonumber \\
&=&q({\bf P}\cdot{\bf H}-{\bf H}\cdot{\bf P}).
 \label{torque}
\end{eqnarray}
Multiplying to the left the expression (\ref{H2}) for $H_{\alpha\gamma}$ by
$P_{\alpha\gamma}=n_\alpha n_\gamma$ and to the right by $n_\gamma
n_\beta$, gives, after some algebra
\begin{eqnarray}
\Gamma \tau_{\alpha\beta}&=& q\Gamma (n_\alpha n_\gamma
H_{\gamma\beta}-H_{\alpha\gamma} n_\gamma n_\beta  ) \nonumber \\
&=& \left[ q^2 (n_\alpha N_\beta-N_\alpha n_\beta)- \frac{\xi q}
{3}(q+2)(n_\alpha n_\gamma
D_{\gamma\beta}-D_{\alpha\gamma}n_\gamma n_\beta )\right].
\label{tau}
\end{eqnarray}
Eq. (\ref{tau}) may now be compared to the antisymmetric part
of Eq. (\ref{ELstress}), to give
\begin{eqnarray}
\alpha_3-\alpha_2 &=& {2 q^2\over \Gamma } = \gamma_1,\\
\alpha_6-\alpha_5 &=& -{2\over 3}q\xi({q+2\over \Gamma}) =
\gamma_2
\end{eqnarray}
where the equalities with $\gamma_1,\gamma_2$ come from comparison
with Eq. (\ref{gammas}). We may slightly rewrite the
antisymmetric term (\ref{tau}) in a form that is closer to the one
used in \cite{EPL}. This can be done by substituting 
the expression for $N_{\mu}$ written in terms of the molecular
field
\begin{equation}
N_{\mu} ={h_{\mu}\over \gamma_1 }-{\gamma_2\over \gamma_1
}n_{\sigma}D_{\sigma \mu}
\label{C6}
\end{equation}
into (\ref{tau}). This gives
\begin{equation}
\Gamma \tau_{\alpha\beta}= \frac{q^2}{\gamma_1} \left
(n_{\alpha}h_{\beta} -h_{\alpha}n_{\beta} \right)
+q^2\frac{\gamma_2}{\gamma_1} \left ( n_{\sigma}D_{\sigma
\alpha}n_{\beta} - n_{\alpha}n_{\sigma}D_{\sigma \beta}\right) -
\frac{\xi q} {3}(q+2)(n_\alpha n_\sigma
D_{\sigma\beta}-D_{\alpha\sigma}n_\sigma n_\beta ).
\end{equation}
Hence the expression for $\tau_{\alpha\beta}$ simplifies to
\begin{equation}
\tau_{\alpha\beta}= \frac{q^2}{\Gamma\gamma_1} \left
(n_{\alpha}h_{\beta} -h_{\alpha}n_{\beta} \right) \label{tau_EPL}
\end{equation} which is the antisymmetric term
in the director field treatment of Ref. \cite{EPL} (see 
eq. (2) of \cite{EPL}).

We now turn to the symmetric part of the total stress tensor
(excluding the active contribution and the double gradient term):
\begin{eqnarray}
\sigma_{\alpha\beta}= &-&P_0 \delta_{\alpha \beta} +2\xi
(Q_{\alpha\beta}+{1\over 3}\delta_{\alpha\beta})Q_{\gamma\epsilon}
H_{\gamma\epsilon}\\\nonumber &-&\xi
H_{\alpha\gamma}(Q_{\gamma\beta}+{1\over
  3}\delta_{\gamma\beta})-\xi (Q_{\alpha\gamma}+{1\over
  3}\delta_{\alpha\gamma})H_{\gamma\beta} 
\label{ssym}
\end{eqnarray}
The active contribution is:
\begin{eqnarray}
\Pi^{\rm active}_{\alpha\beta}=
-\zeta q n_\alpha n_\beta+\zeta\frac{q}{3}
\delta_{\alpha\beta}.
 \label{ssyma}
\end{eqnarray}
Note that the double gradient term term $-\partial_\alpha
Q_{\gamma\nu} {\delta {\cal F}\over \delta\partial_\beta
Q_{\gamma\nu}}$ is analogous to the director field
term $-\partial_\alpha
n_{\nu} {\delta {\cal F}\over \delta\partial_\beta
n_{\nu}}$, which is not included in Eq. (\ref{ELstress}) hence
not considered hereafter.

By using Eq. (\ref{H2}) for ${\bf H}$,
after some algebra, one obtains the complete
expression for $\sigma_{\alpha\beta}$ as
\begin{eqnarray}
\sigma_{\alpha\beta}&=& -\frac{q\xi}{3\Gamma}(q+2) (n_\beta
N_\alpha+n_\alpha N_\beta)+ \frac{\xi^2 q} {3\Gamma}(4-q)\left
(D_{\alpha\gamma}n_\gamma n_\beta+ n_\alpha n_\gamma
D_{\gamma\beta}\right) \nonumber \\
&+& \frac{4}{9\Gamma}(q-1)^2 \xi D_{\alpha\beta}
+ \frac{2}{3\Gamma}q^2 \xi^2 (4q^2-4q-3)n_\alpha
n_\beta D_{\gamma\nu}n_\nu n_\gamma\nonumber \\
&+&\frac{q\xi^2}{\Gamma}(4-7q-8q^2+8q^3) \delta_{\alpha\beta}D_{\gamma\nu}n_\nu
n_\gamma \nonumber 
\label{ssym3}
\end{eqnarray}

The first term of the right hand side of Eq. (\ref{ssym3}) can be
usefully rewritten (for comparison with the equation in \cite{EPL}) 
by using (\ref{C6}) to write $N_\mu$ in terms of
$h_\mu$. 
\begin{eqnarray}
-q\xi/3 (q+2) (n_\beta N_\alpha+n_\alpha N_\beta)&=& -q\xi/3
(q+2)\left ( n_\beta\frac{h_\alpha}{\gamma_1}-
n_\beta\frac{\gamma_2}{\gamma_1}n_\sigma D_ {\sigma\alpha} +
n_\alpha\frac{h_\beta}{\gamma_1} -
n_\alpha\frac{\gamma_2}{\gamma_1}n_\sigma D_
{\sigma\beta}\right)\nonumber\\
&=&\frac{\nu_{EL}\Gamma}{2}\left ( n_\beta h_\alpha+ n_\alpha
h_\beta\right) -\frac{\nu_{EL}\Gamma\gamma_2}{2}\left (n_\beta
n_\sigma D_ {\sigma\alpha} + n_\alpha n_\sigma D_
{\sigma\beta}\right)
\end{eqnarray}
where in the last line we have used Eq. (\ref{nuEPL}).

The Navier Stokes equation in the Stokes regime  is
\begin{eqnarray}\label{navierstokes_stokes}
\eta \partial_\beta(\partial_\alpha
u_\beta + \partial_\beta u_\alpha) = 
2\partial_{\beta}\eta D_{\alpha \beta} =
- \partial_\beta (\Pi_{\alpha\beta}).
\end{eqnarray}
$-\Pi_{\alpha\beta}$ can equivalently be rewritten as
\begin{eqnarray}
\label{navierstokes3}
-\Pi_{\alpha \beta} &=& 
-\frac{\nu_{EL}}{2}\left ( n_\beta h_\alpha+ n_\alpha
h_\beta\right)+ \frac{4\xi^2 } {9\Gamma}(q-1)^2\left
(D_{\alpha\sigma}n_\sigma n_\beta+ n_\alpha n_\sigma
D_{\sigma\beta}\right) \nonumber \\
&-& \frac{4}{9\Gamma}(q-1)^2 \xi^2 D_{\alpha\beta} 
- \frac{2}{3\Gamma}q^2 \xi^2
(4q^2-4q-3)n_\alpha
n_\beta D_{\gamma\nu}n_\nu n_\gamma\nonumber \\
&-&\frac{q\xi^2}{\Gamma}(4-7q-8q^2+8q^3)
\delta_{\alpha\beta}D_{\gamma\nu}n_\nu n_\gamma \nonumber\\
&+&
\zeta q n_\alpha
n_\beta - \zeta\frac{q}{3}\delta_{\alpha\beta} \nonumber \\
&-&
\frac{q^2}{\Gamma\gamma_1} \left
(n_{\alpha}h_{\beta} -h_{\alpha}n_{\beta} \right).
\end{eqnarray}

If $\lambda=\zeta=0$, i.e. for passive liquid
crystals, Eq. (\ref{ssym3}) gives the symmetric part of the 
Beris-Edwards stress
(ignoring the distortion stress) and this, together with Eq.
(\ref{tau}) gives the Leslie coefficients which are listed
in Section III B (Eqs. \ref{viscosity1}--\ref{viscosity6}).

In Eq. (\ref{navierstokes3}) the term proportional to $D_{\alpha\beta}$
may be added to the left hand side in Eq. \ref{navierstokes_stokes}
to renormalise the apparent viscosity, while the rest of it
may be rewritten as 
\begin{eqnarray}
\label{navierstokes4}
&-& \frac{\nu_{EL}}{2}\left ( n_\beta
h_\alpha+ n_\alpha h_\beta\right)+ \frac{4\xi^2 }
{9\Gamma}(q-1)^2\left (D_{\alpha\sigma}n_\sigma n_\beta+ n_\alpha
n_\sigma
D_{\sigma\beta}\right) \nonumber \\
&-& \frac{2}{3\Gamma}q^2 \xi^2 (4q^2-4q-3)n_\alpha
n_\beta D_{\gamma\nu}n_\nu n_\gamma\nonumber \\
&-&\frac{q\xi^2}{\Gamma}(4-7q-8q^2+8q^3)
\delta_{\alpha\beta}D_{\gamma\nu}n_\nu n_\gamma \nonumber\\
&+&
\zeta q n_\alpha
n_\beta - \zeta\frac{q}{3}\delta_{\alpha\beta} \nonumber \\
&-& \frac{1}{2} \left (n_{\alpha}h_{\beta} -h_{\alpha}n_{\beta}
\right)
\end{eqnarray}
where for the last term we have used relation (\ref{gammas}). By
comparing our equation with the one in \cite{EPL} we then get
Eqs. (\ref{etaEPL}), (\ref{nuEPL}) in the text.

\newpage

\label{ordersec}

\begin{thebibliography}{99}
\bibitem{ramaswamy} Y. Hatwalne, S. Ramaswamy, M. Rao, R. A. Simha,
{\it Phys. Rev. Lett.} {\bf 92}, 118101 (2004).
\bibitem{ramaswamy2} R.A. Simha, S. Ramaswamy,
{\it Phys. Rev. Lett.} {\bf 89}, 058101 (2002).
\bibitem{kruse1} K. Kruse, J. F. Joanny, F. Julicher, J. Prost,
K. Sekimoto, {\it Eur. Phys. J. E} {\bf 16}, 5 (2005).
\bibitem{kruse2} K. Kruse {\it et al.}, {\it Phys. Rev. Lett.}
{\bf 92}, 078101 (2004).
\bibitem{joanny} R. Voituriez, J. F. Joanny, J. Prost,
{\it Phys. Rev. Lett.} {\bf 96}, 028102 (2006).
\bibitem{liverpool} T. B. Liverpool, M. C. Marchetti
{\it Phys. Rev. Lett.} {\bf 90}, 138102 (2001); {\it Europhys. Lett.}
{\bf 69}, 846 (2005); cond-mat/0607285.
\bibitem{EPL} R. Voituriez, J. F. Joanny, J. Prost,
{\it Europhys. Lett.} {\bf 70}, 404 (2005).
\bibitem{ramaswamy3} V. Narayan, N. Menon, S. Ramaswamy,
{\it J. Stat. Mech.: Theory and Experiment} P01005 (2006).
\bibitem{llopis} I. Llopis, I. Pagonabarraga, {\it Europhys. Lett.},
{\bf 75}, 999 (2006).
\bibitem{ignacio} S. Ramachandran, P. B. S. Kumar, I. Pagonabarraga,
{\it Eur. Phys. J. E} {\bf 20}, 151 (2006).
\bibitem{bray}  D. Bray, {\em Cell movements: from molecules to
motility}, Garland Publishing, New York (2000).
\bibitem{peter} P. R. Cook, {\it Principles of Nuclear Structure and
Function}, Wiley, New York (2001).
\bibitem{beads} J. van der Gucht, E. Paluch, C. Sykes,
{\it Proc. Natl. Acad. Sci. USA} {\bf 102}, 7847 (2005);
M. F. Carlier {\it et al.}, {\it Bioessays} {\bf 25}, 336 (2003);
N. J. Burroughs, D. Marenduzzo, {\it Phys. Rev. Lett.}
{\bf 98}, 238302 (2007).
\bibitem{cytoskeleton} J. Howard, {\it Mechanics of Motor Proteins
and the Cytoskeleton}, Sinauer Associates, Inc., Sunderland (2001).
\bibitem{active_actin_myosin} D. Humphrey {\it et al.},
{\em Nature} {\bf 416}, 413 (2002).
\bibitem{surrey} T. Surrey, F. Nedelec, S. Leibler and E. Karsenti, 
{\it Science} {\bf 292}, 1167 (2001).
\bibitem{nedelec} F. J. Nedelec {\it et al.}, {\it Nature} {\bf 389},
305 (1997).
\bibitem{kierfeld} P. Kraikivski, R. Lipowsky, J. Kierfeld,
{\it Phys. Rev. Lett.} {\bf 96}, 258103 (2006).
\bibitem{lubensky}  C. Storm, J.J. Pastore, F.C. MacKintosh, 
T.C. Lubensky, P.A. Janmey, {\it Nature} {\bf 435}, 191 (2005).
\bibitem{activeLB} D. Marenduzzo, E. Orlandini, M. E. Cates,
J. M. Yeomans, {\it J. Non-Newt. Fluid Mech.},
in press (2007); D. Marenduzzo, E. Orlandini,
J. M. Yeomans, {\it Phys. Rev. Lett.} {\bf 98}, 118102 (2007).
\bibitem{nomenclature} {An active particle ``absorbs energy from its 
surroundings and dissipates it in the process of carrying out internal 
movements, usually resulting in translatory or rotary motion'' 
\cite{ramaswamy}.}
\bibitem{note_noneq} {As is conventional \cite{chem_noneq_book}, we use
``phase transition'' to refer to a 
singular dependence of observable quantities on model parameters
even in a non-equilibrium system, for which
there is no underlying thermodynamic free energy to lead to distinct
equilibrium ``phases''. }
\bibitem{chem_noneq_book} H. Haken, {\it Synergetics: An Introduction.
Nonequilibrium Phase Transitions and Self-Organization in Physics, Chemistry
and Biology}, 3rd rev. enl. ed. New York: Springer-Verlag (1983).
\bibitem{forest} M. G. Forest, R. H. Zhou, Q. Wang,
{\it Phys. Rev. Lett.} {\bf 93}, 088301 (2004).
\bibitem{ramaswamy_chaos}  B. Chakrabarti, M. Das, C. Dasgupta, 
S. Ramaswamy, A. K. Sood, {\it Phys. Rev. Lett.} {\bf 92}, 055501
(2004).
\bibitem{mike} A. Aradian, M.E. Cates, {\it Europhys. Lett.}
{\bf 70}, 397 (2005).
\bibitem{alexander} A. N. Morozov, W. van Saarloos,
{\it Phys. Rev. Lett.} {\bf 95}, 024501 (2005).
\bibitem{onuki1} A. Onuki, {\em Phase Transition Dynamics}, 
Cambridge University Press, Cambridge (2002).
\bibitem{onuki2} A. Onuki, {\it J. Phys.: Condens. Matt.} {\bf 10},
49, 11473 (1998).
\bibitem{chem_noneq_review} M. C. Cross, P. C. Hohenberg,
{\it Rev. Mod. Phys.} {\bf 65}, 851 (1993).
\bibitem{muthukumar} S. C. Glotzer, A. E. Di Marzio, 
M. Muthukumar, {\it Phys. Rev. Lett.} {\bf 74}, 2034 (1995).
\bibitem{degennes} P.G. de Gennes and J. Prost, {\it The Physics of
Liquid Crystals, 2nd Ed.}, Clarendon Press, Oxford, (1993).
\bibitem{beris}
A.N. Beris and B.J. Edwards, {\it Thermodynamics of Flowing Systems},
Oxford University Press, Oxford, (1994); A.N. Beris, B.J. Edwards and
M. Grmela, {\it J. Non-Newtonian Fluid Mechanics}, {\bf 35} 51 (1990). 
\bibitem{O92}
P.D. Olmsted and P.M. Goldbart, Phys. Rev. A {\bf 46}, 4966 (1992).
\bibitem{O99}
P.D. Olmsted and C.-Y. David Lu, Phys. Rev. E {\bf 56}, 55 (1997);
ibid, {\bf 60}, 4397 (1999).
\bibitem{thoumine} O. Thoumine, A. Ott, {\it J. Cell. Sci.}
{\bf 110}, 2109 (1997).
\bibitem{frey} J. Uhde, M. Keller, E. Sackmann, A. Parmegiani,
E. Frey, {\it Phys. Rev. Lett.} {\bf 93}, 268101 (2004).
\bibitem{succi} S. Succi, {\it The Lattice Boltzmann equation}, 
Oxford University Press (2001).
\bibitem{colin} C. Denniston, E. Orlandini, J. M. Yeomans,
{\it Phys. Rev. E} {\bf 63}, 056702 (2001).
\bibitem{lblc} C. Denniston, D. Marenduzzo, E. Orlandini, 
J. M. Yeomans, {\it Phil. Trans. R. Soc. Lond. A} {\bf 362}, 1745 (2004).
\bibitem{rheoHAN} D. Marenduzzo, E. Orlandini, J. M. Yeomans,
{\it Europhys. Lett.} {\bf 64}, 406 (2003); {\it J. Chem. Phys.}
{\bf 121}, 582 (2004).
\bibitem{active_chiral_gel} D. Marenduzzo, S. Ramaswamy, M. E. Cates, 
work in progress.
\bibitem{nidhal} N. Sulaiman, D. Marenduzzo, J.M. Yeomans, 
{\it Phys. Rev. E} {\bf 74}, 041708 (2006).
\end{thebibliography}
\end{document}